\documentclass[twocolumn, nofootinbib, aps, prl, floats, floatfix, amsmath, amssymb, longbibliography, superscriptaddress, preprintnumbers]{revtex4-1} %
\usepackage[final]{graphicx}
\usepackage{hyperref}
\usepackage{amsmath}
\usepackage{bbm}
\usepackage{bm}
\usepackage{amsfonts}
\usepackage{amssymb}
\usepackage{latexsym}
\usepackage{graphicx}
\usepackage[english]{babel}
\usepackage{multirow}
\usepackage{float}
\usepackage{url}
\usepackage{slashed}
\usepackage{xcolor} 
\usepackage[utf8]{inputenc}
\usepackage{stmaryrd} 
\usepackage{enumitem}
\usepackage{hyperref}
\usepackage{cleveref}
\usepackage{siunitx}
\usepackage{verbatim}
\usepackage{braket}
\usepackage{mathrsfs}

\usepackage{amsthm}

\theoremstyle{remark}

\newcommand{\be}{\begin{equation}}
\newcommand{\ee}{\end{equation}}
\newcommand{\ba}{\begin{array}}
\newcommand{\ea}{\end{array}}
\newcommand{\bea}{\begin{eqnarray}}
\newcommand{\eea}{\end{eqnarray}}

\newcommand{\nn}{\nonumber}

\newcommand{\bk}{{\bf k}}
\newcommand{\bp}{{\bf p}}

\newcommand{\bx}{{\bf x}}
\newcommand{\bq}{{\bf q}}

\newcommand{\besub}{\begin{subequations}}
\newcommand{\eesub}{\end{subequations}}

\newcommand{\nnn}{\nonumber \\}
\newcommand{\mkz}{m_{\rm KZ}}
\newcommand{\I}{\mathrm{i}}
\newcommand{\CX}{\mathbf{X}}

\newcommand{\beq}{\begin{equation} \begin{aligned}}
\newcommand{\eeq}{\end{aligned} \end{equation}}
\newcommand{\ts}{\tilde\sigma}
\newcommand{\tT}{\tilde T^{TT}}
\newcommand{\sh}{{\tilde h}}

\definecolor{darkerblue}{rgb}{0.2,0.2,0.5}
\definecolor{seagreen}{rgb}{0.180392,0.545098,0.341176}
\definecolor{smagenta}{rgb}{0.5,0.145098,0.341176}
\definecolor{deepblue}{rgb}{0,0,1}

\begin{document}

	\title{Gravitational Waves Produced by Domain Walls During Inflation}
	
	\author{Haipeng An}
	\email{anhp@mail.tsinghua.edu.cn}
	\affiliation{Department of Physics, Tsinghua University, Beijing 100084, China}
	\affiliation{Center for High Energy Physics, Tsinghua University, Beijing 100084, China}
	\affiliation{Center for High Energy Physics, Peking University, Beijing 100871, China}	
	\affiliation{Frontier Science Center for Quantum Information, Beijing 100084, China}
	
	\author{Chen Yang}
	\email{yangc18@mails.tsinghua.edu.cn}
	\affiliation{Department of Physics, Tsinghua University, Beijing 100084, China}

\begin{abstract}
	
We study the properties of the stochastic gravitational wave background (SGWB) produced by domain walls (DWs) during inflation without forming a network. We numerically simulate the DW production caused by a second-order phase transition and calculate the SGWB spectrum using a $1000\times1000\times1000$ lattice. We show that the SGWB can be observed directly by future terrestrial and spatial gravitational wave detectors and through the B-mode spectrum in CMB. This signal can also explain the common noise process observed by pulsar timing array experiments. With numerical simulations, we derive an empirical formula for the strength and qualitative features of the SGWB spectrum. The details of the SGWB spectrum also contain information about the later evolution of the universe.

\end{abstract}
\maketitle

\section{Introduction}
	
The direct discovery of gravitational waves (GWs) produced by black hole binary merger opens up a new era of GW astrophysics~\cite{LIGOScientific:2016aoc}. GWs can also be produced in the early universe via phase transitions, topological defects, scalar perturbations, and quantum fluctuation during inflation, forming the stochastic GW background. See Ref.~\cite{Caldwell:2022qsj} for an excellent review and references therein. GWs, once produced, propagate almost freely through the universe, bringing us information about their origin and the evolution of the universe. 
Terrestrial and space-based detectors are proposed to search for SGWB~\cite{Seoane:2013qna,Audley:2017drz,Kawamura:2011zz,Luo:2015ght,Guo:2018npi,Crowder:2005nr,Harry:2006fi,Corbin:2005ny,Kramer:2013kea,Abramovici:1992ah,TheVirgo:2014hva,Punturo:2010zz,Reitze:2019iox}. SGWB with a frequency around $10^{-8}$ Hz can be detected via radio telescopes using the pulsar timing arrays (PTAs)~\cite{Hobbs:2009yy,Kramer:2013kea,Janssen:2014dka}. The common noise process observed by the PTAs might indicate a signal of SGWB~\cite{NANOGrav:2020bcs,Goncharov:2021oub,Li:2017drr,Antoniadis:2022pcn}. SGWB with longer wavelengths may also leave prints on the cosmic microwave background and be detected via the B-mode spectrum~\cite{Array:2015xqh,Hui:2018cvg,Abazajian:2019eic}

%

One important source of SGWB that has been extensively studied in the literature is the cosmic DWs~\cite{Zeldovich:1974uw,Vilenkin:1981zs,Gelmini:1988sf,Coulson:1995nv,Hiramatsu:2012sc,Hiramatsu:2013qaa,Garcia-Bellido:2007fiu,Krajewski:2021jje} (for a review, see \cite{Saikawa:2017hiv}). DW-induced GWs as a test for new physics beyond the standard model have attracted a lot of interests in the literature~\cite{Dine:2010eb,Chen:2020wvu,Chen:2020soj,Zhou:2020ojf,Blasi:2022woz,Barman:2022yos,Gelmini:2020bqg,Gelmini:2021yzu,Wu:2022tpe}. DWs are 2D topological defects created via the spontaneous breaking of discrete symmetries. 
Stable domain walls will form a network in the radiation domination (RD) era and overclose the universe~\cite{Zeldovich:1974uw}. Thus an explicit symmetry broken term is usually introduced to ensure the DWs annihilate before the big bang nuclear synthesis~\cite{Vilenkin:1981zs,Gelmini:1988sf,Coulson:1995nv}. 
Traditional studies of SGWB produced by DWs focus on production through the DW network. Both qualitative analysis~\cite{Hiramatsu:2012sc} and numerical simulations~\cite{Hiramatsu:2013qaa} show that the GW energy density production rate by the DW network is constant during RD. Therefore the GWs produced right before the annihilation of the DWs dominate the contribution to the SGWB energy density today since they are the least redshifted. However, if the DWs annihilate too early, the SGWB signal will be too weak to induce detectable signals in future GW detectors. 

In this letter, we point out that if the DWs were produced during inflation, they would generate SGWB during inflation without forming a network, and the strength of the SGWB can reach the sensitivities of the planned GW detectors. In flat space-time or during the RD era, static sources cannot radiate GWs due to energy conservation. However, during inflation, the Hubble expansion rate is significant, and thus energy is no longer conserved. Therefore even {\it comovingly static} sources can produce GWs. If the DWs are produced about 60 e-folds before the end of inflation, the SGWB will induce a CMB B-mode spectrum
that future CMB observers can detect. If the phase transition happens at about 40 e-folds before the end of inflation, the SGWB the DWs produced will be able to explain the common noise process observed by PTAs~\cite{NANOGrav:2020bcs,Goncharov:2021oub,Li:2017drr,Antoniadis:2022pcn}.


\bigskip

\section{DW production during inflation}


This work studies the phase transition in a spectator sector during de Sitter inflation. With conformal time $\tau$, the scale factor is $a(\tau) = - 1/H\tau$. 
%
%
%
We assume the spectator sector is in the Landau-Ginzburg type
\bea\label{eq:1}
V = - \frac{1}{2} m_{\rm eff}^2 \sigma^2 + \frac{\lambda}{4} \sigma^4 \ ,
\eea 
where $\sigma$ is the spectator field and serves as the order parameter for the second-order phase transition. 

We consider two scenarios. 
In scenario (A), we assume the universe was in RD before inflation. At the beginning of inflation, the temperature of the thermal plasma quickly falls, and thus phase transition happened in the plasma. This scenario has been considered in the context in of grand unification models~\cite{Jiang:2015qor}. In this case, we have $m_{\rm eff}^2 = y T^2 - m^2$, where $T$ is the temperature in the spectator sector, $m$ is the mass of $\sigma$ at zero temperature, and $y$ is a constant parameter. In this scenario, phase transition happens at the critical temperature $T_c = y^{-1/2} m$.
%
%
In scenario (B), we consider the inflaton-driven phase transition similar to the study for first-order phase transitions in Refs.~\cite{An:2020fff,An:2022cce}. We observe that in large field inflation models, the excursion of the inflaton field can be as large as the Planck scale. Thus, if we assume that the spectator fields are directly coupled to the inflaton field, the evolution of the inflaton field may induce dramatic changes of the masses or couplings in the spectator sector and thus leads to phase transitions during inflation.   
The effective mass can be written as $m_{\rm eff}^2 = y \phi^2 - m^2$, where $\phi$ is the inflaton field. 

In this work, we use lattice simulation to numerically study the evolution of the spectator field $\sigma$ after the phase transition. The details of the simulation method are presented in the appendix. Here we describe the qualitative features of evolution and the DW formation process.  

The production of the topological defects can be described by the Kibble-Zurek mechanism~\cite{Kibble:1976sj,Zurek:1985qw}. Around the critical time $\tau_c$, we can use the potential
\bea
V_{\rm KZ} = - \frac{1}{2} \mkz^3 a_c^{-1} (\tau - \tau_c) \sigma^2 + \frac{\lambda}{4} \sigma^4 \ ,
\eea
to describe the evolution of $\sigma$,
where $a_c \equiv a_{\tau=\tau_c}$. For scenarios (A) and (B), the Kibble-Zurek scale $\mkz$ can be written as~\cite{Murayama:2009nj}
\bea
m_{{\rm KZ}(A)}^3 &=& - y a_c \left.\frac{d T^2}{d \tau}\right|_{\tau=\tau_c} = 2 m^2 H \ , \nnn 
m_{{\rm KZ}(B)}^3 &=& - y a_c \frac{d\phi^2_0}{d\tau} = \frac{2^{3/2} \varepsilon^{1/2} m^2 H M_{\rm pl}}{\phi_0(\tau_c)} \ ,
\eea 
where $\epsilon$ is the slow-roll parameter, and $\phi_0$ is the homogeneous part of $\phi$. In the simplest slow-roll model, $\phi_0(\tau_c) \approx N_c (2\varepsilon)^{1/2} M_{\rm pl}$~\cite{Baumann:2009ds}, with $N_c$ the e-folds between the critical point and the end of inflation. Therefore, compared to (A), $\mkz$ in (B) is suppressed by a factor of $N_c^{1/3}$. In this work, for the phase transition to complete during inflation, we assume $m$ is large enough, such that $\mkz \gg H$. 

Right after the phase transition, the evolution of the Fourier modes of the $\sigma$ field is driven by the linearized equation of motion
\bea\label{eq:sigmak}
\sigma_\bk'' + \frac{2 a'}{a} \sigma'_\bk + \omega^2_\bk(\tau) \sigma_\bk = 0 \ ,
\eea
where the primes denote the derivatives to $\tau$, and the temporary frequency $\omega_\bk(\tau)$ is given by 
\bea\label{eq:omegak}
\omega_\bk^2 = k^2 - a_c^2 m_{\rm KZ}^3(\tau - \tau_c) + \frac{\lambda}{2} \langle \sigma^2(\tau,\bx)\rangle \ ,
\eea
where $\langle \sigma^2(\tau,\bx) \rangle$ is the temporary expectation value of $\sigma^2(\tau,\bx)$. 
Around $\tau_c$, $\sigma$ is still located at 0; thus $\langle\sigma^2(\tau)\rangle$ can be neglected. Therefore $\omega_\bk^2$ is negative for the infrared modes with small enough $k^2$, and thus $\sigma_\bk$ grows exponentially. After a while, when the nonlinear term in the potential becomes important, the exponential growth stops. Our numerical simulation shows that only modes with $k a^{-1}(\tau_c) < {\cal O}(\mkz)$ can experience exponential growth. 
Then we follow the strategy in Ref.~\cite{Garcia-Bellido:2002fsq} to simulate the evolution of the $\sigma$ field after the phase transition. 
Due to the exponential growth, the particle numbers for long-wavelength modes are large. Thus we can use the classical Gaussian random field to approximate the quantum modes to set the initial condition for the nonlinear evolution. 
Then we use a $1000\times1000\times1000$ lattice to simulate the nonlinear evolution classically. The details of the simulation are presented in the appendix. 

Fig.~\ref{fig:sigma} shows the evolution of the temporary root-mean-square of $\sigma$ for scenarios (A) and (B) together with the temporary vacuum expectation values $v_{\rm temp}\equiv\lambda^{-1/2}m_{\rm eff}$. The curves for $\sigma_{\rm rms}$ and $v_{\rm temp}$ are normalized to $v = \lambda^{-1/2}m$. One can see that in (A), since the temperature $T$ redshifts as $a^{-1}$, $v_{\rm temp}$ approaches $v$ after about one e-fold. However, in (B), since the evolution is driven by $\phi_0$, $v_{\rm temp}$ will not reach $v$ until the end of inflation. At the beginning of the phase transition, the growth of $\sigma_{\rm rms}$ lags behind $v_{\rm temp}$, which causes the oscillation of $\sigma_{\rm rms}$ around. As a result, $\sigma_{\rm rms}$ oscillates around $v_{\rm temp}$ in the later oscillation. However, the oscillation is significantly damped after several e-folds due to the Hubble friction, and then the $\sigma$ field configuration becomes {\it comovingly static}. 
%
The configurations for $\sigma$ at different times are shown in Figs.~\ref{fig:configA} and \ref{fig:configB}. In both cases, the field configuration becomes {\it static} after several e-folds. In Fig.~\ref{fig:configA}, the $x$ and $y$ axes are the comoving coordinates in the unit of the comoving Kibble-Zurek length, $\mkz^{-1}a_c^{-1}$. One can see that once the DWs are formed, their comoving density does not change, and the average distance between the walls is about $2\pi \mkz^{-1} a_c^{-1}$, as predicted, since only modes with $k a_c^{-1} < {\cal O}(\mkz)$ can grow exponentially.


\begin{figure}[htpb]
	\centering
	\includegraphics[width=0.8\linewidth]{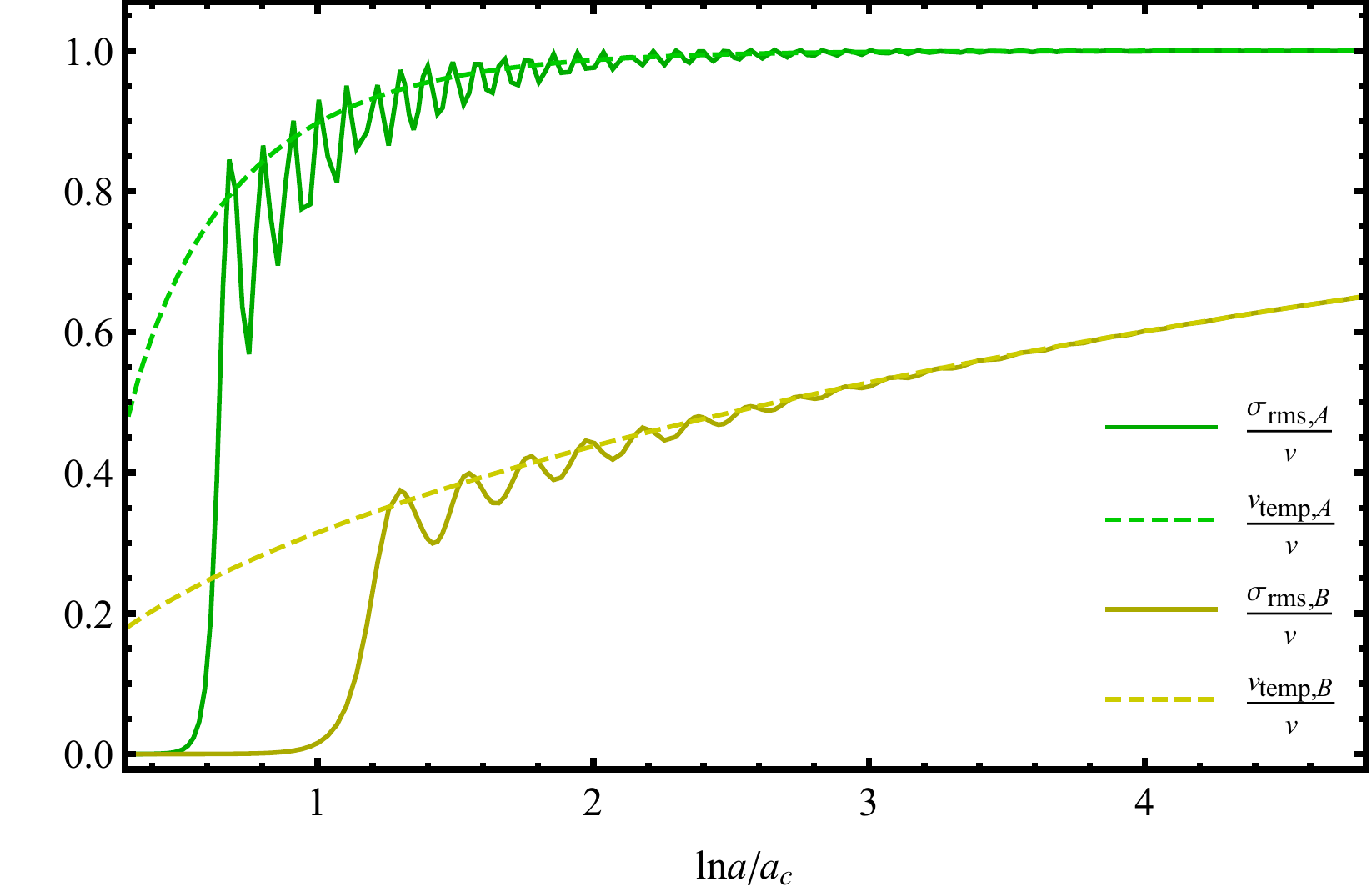}
	\caption{The evolutions of the root mean square of $\sigma$ for scenario (A) (green) and (B) (yellow). The dashed curves show the corresponding the temporary expectation of $\sigma$. For both cases we use $H = 10^{-4}M_{\rm pl}$, $m = 5\times10^{-3} M_{\rm pl}$. For (A) we have $\lambda = 0.0625$, $\mkz = (2 m^2 H)^{1/3}$, and for (B) $\lambda = 0.004$ and $\mkz = 0.126m$. }\label{fig:sigma}
\end{figure}

\begin{figure}[htpb]
	\centering
	\includegraphics[width=0.45\linewidth]{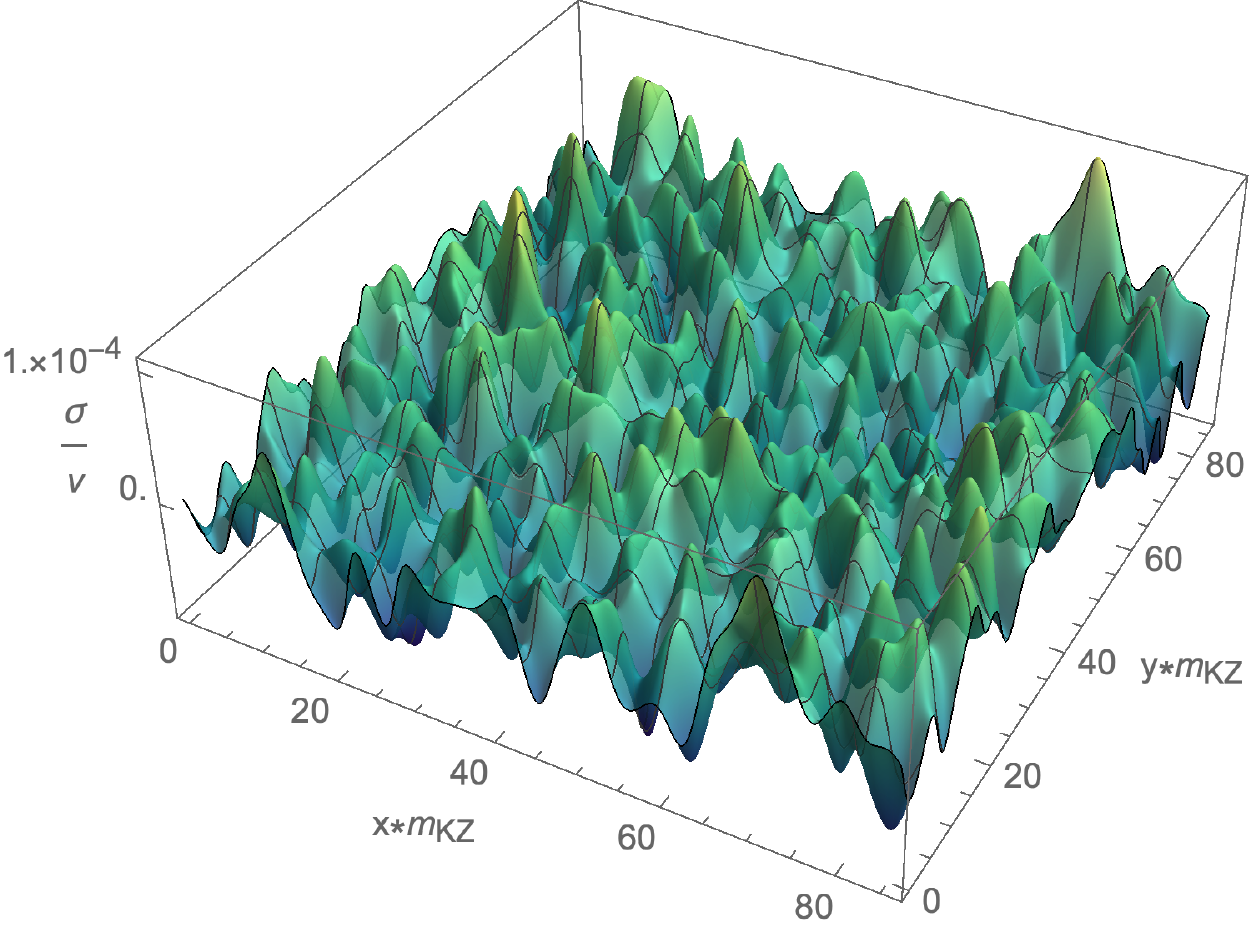}
	\includegraphics[width=0.45\linewidth]{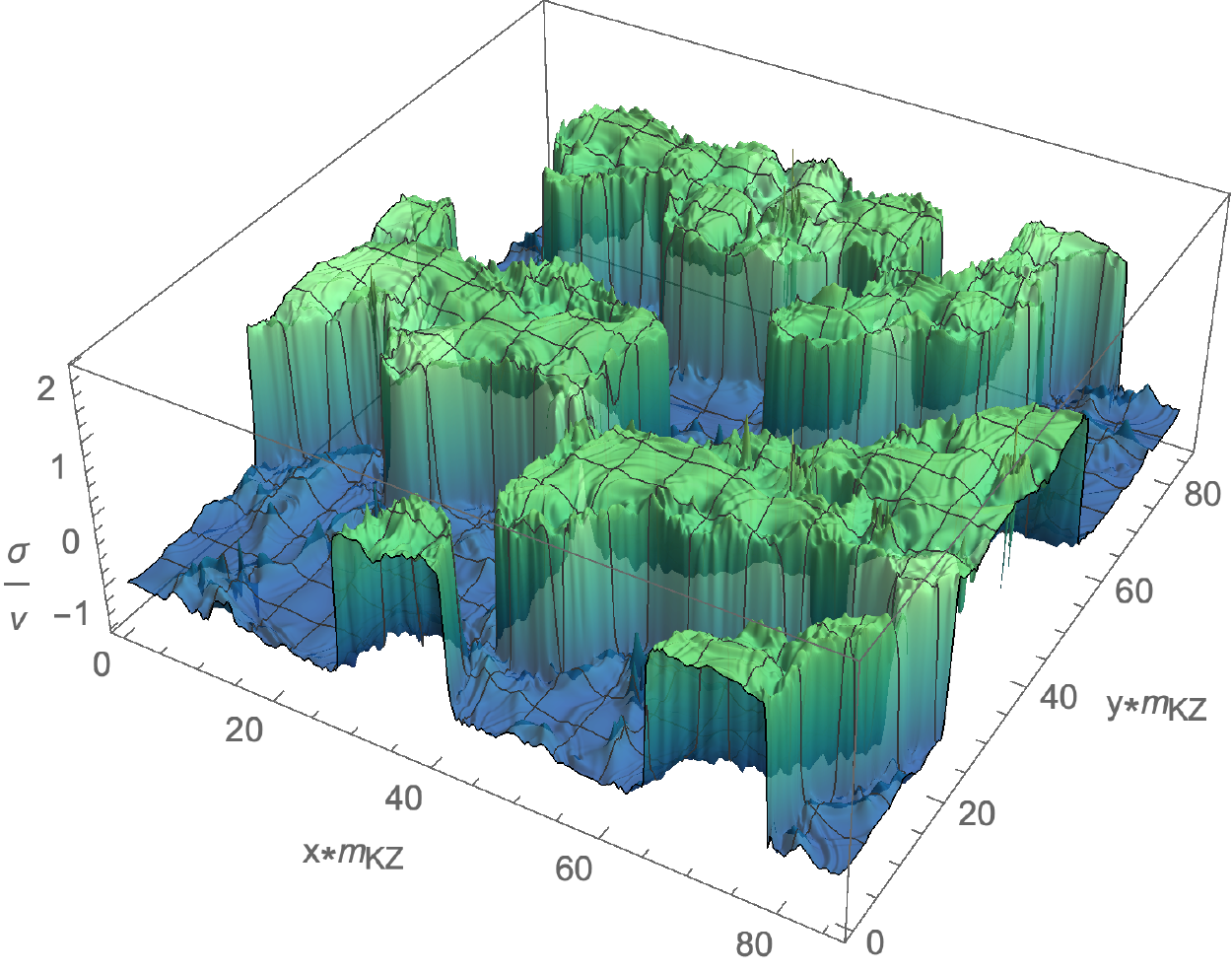}
	\includegraphics[width=0.45\linewidth]{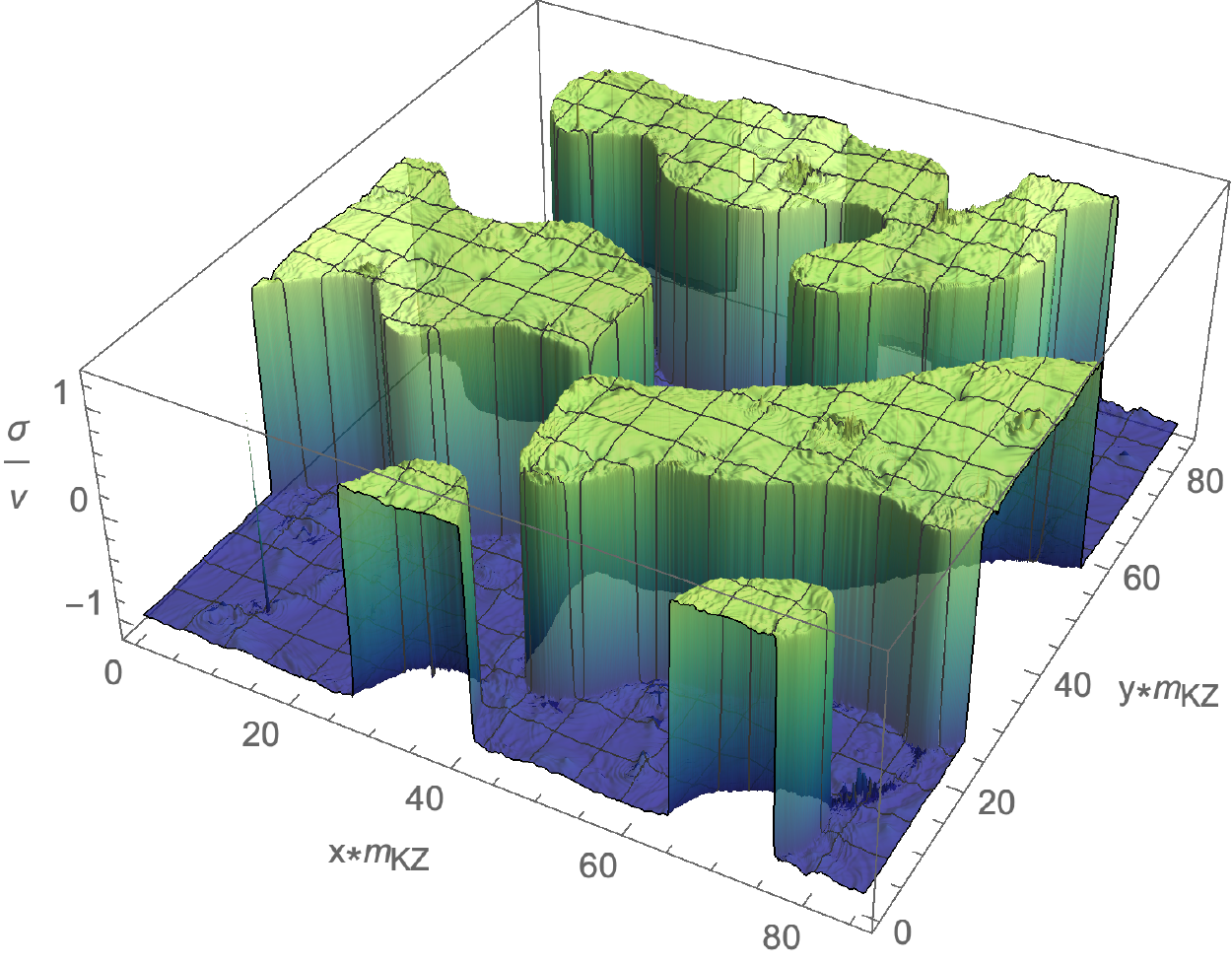}
	\includegraphics[width=0.45\linewidth]{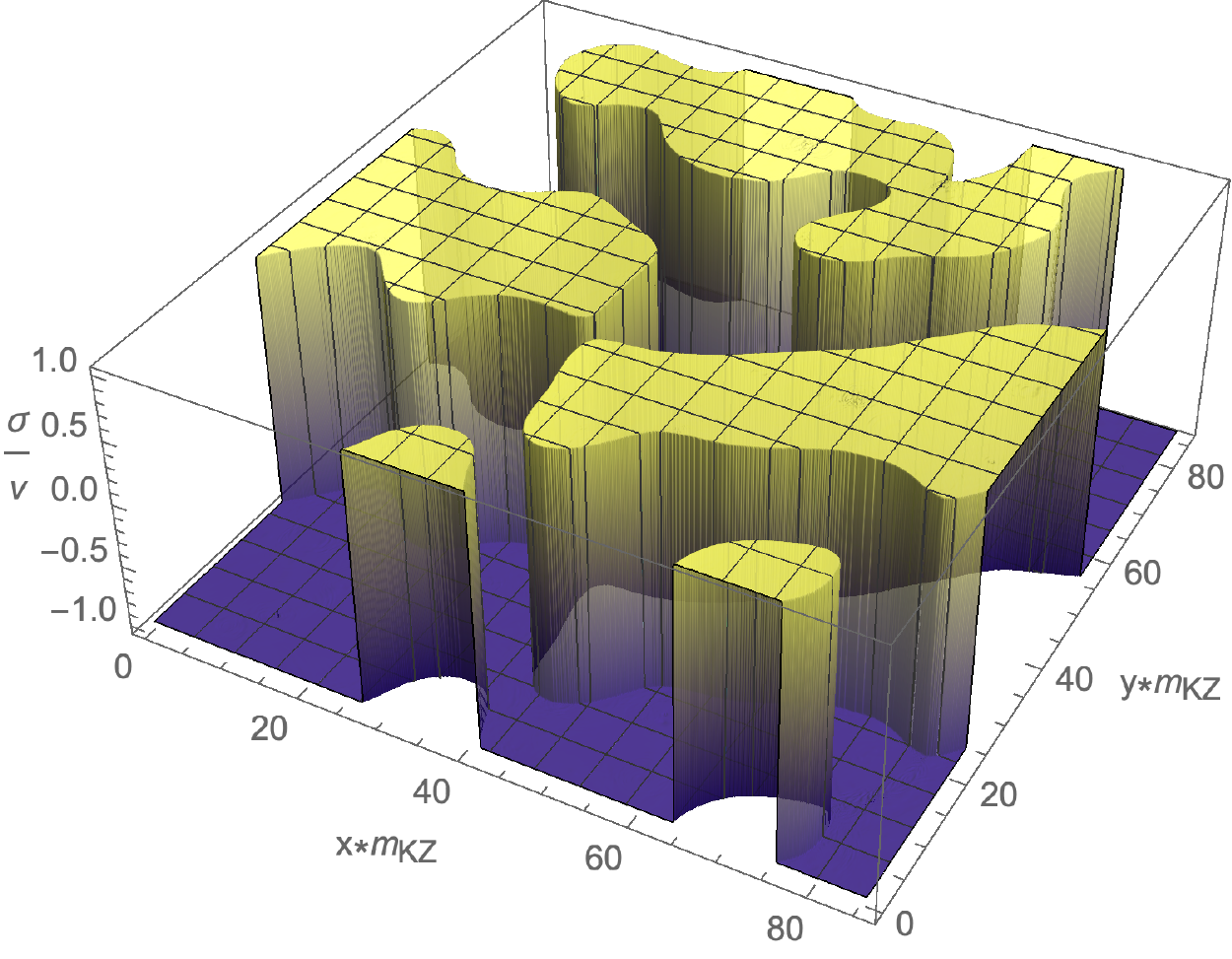}
	\caption{Evolution of the $\sigma$ field in scenario (A) for $\ln a/a_c = 1.3, 2.3, 3.3$ and $4.3$, where the $x$ and $y$ axes show the comoving coordinates in the unit of $a_c^{-1} \mkz^{-1}$, and the $z$ axis shows $\sigma/v$. The parameters in the potential are the same as in Fig.~\ref{fig:sigma}. }\label{fig:configA}
\end{figure}

\begin{figure}[htpb]
	\centering
	\includegraphics[width=0.45\linewidth]{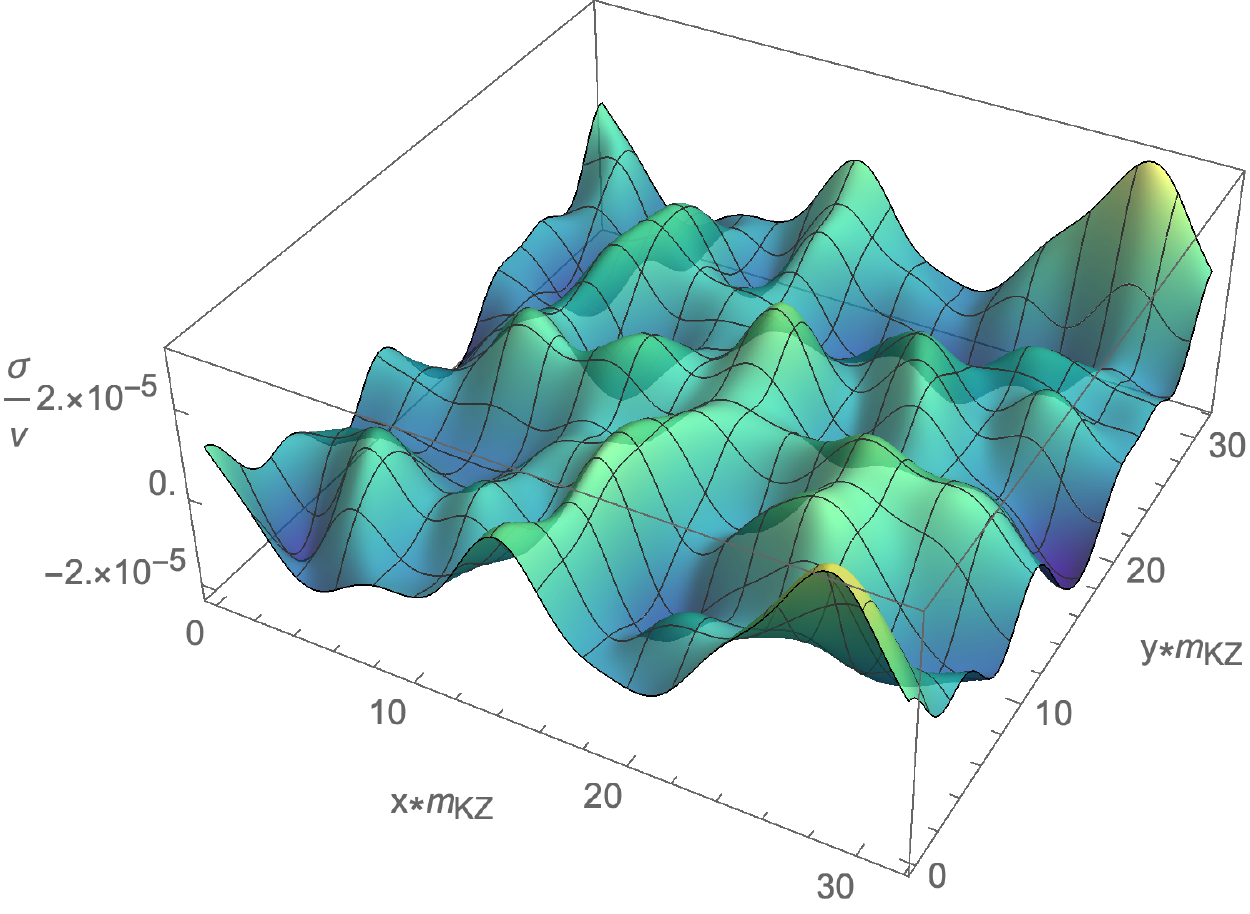}
	\includegraphics[width=0.45\linewidth]{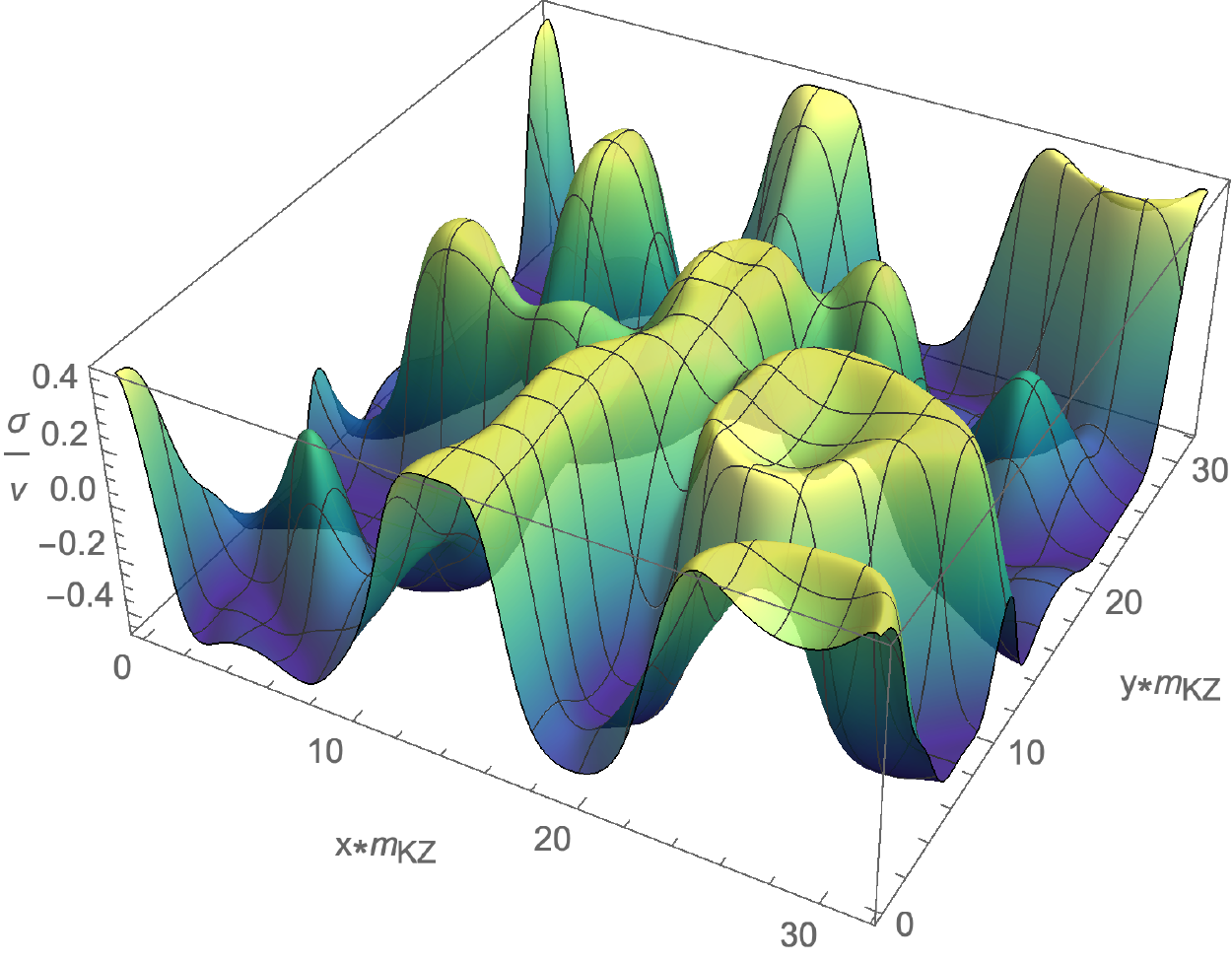}
	\includegraphics[width=0.45\linewidth]{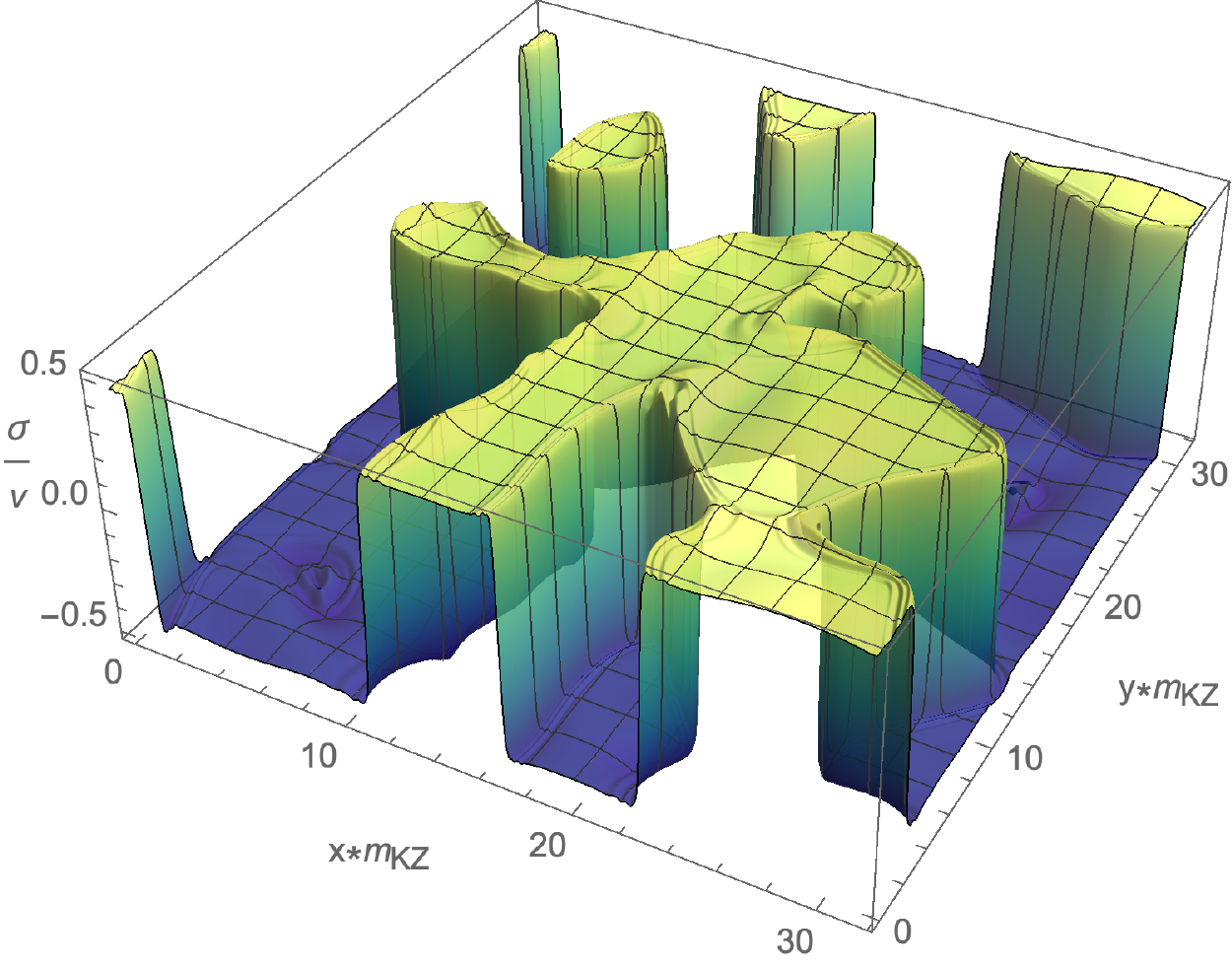}
	\includegraphics[width=0.45\linewidth]{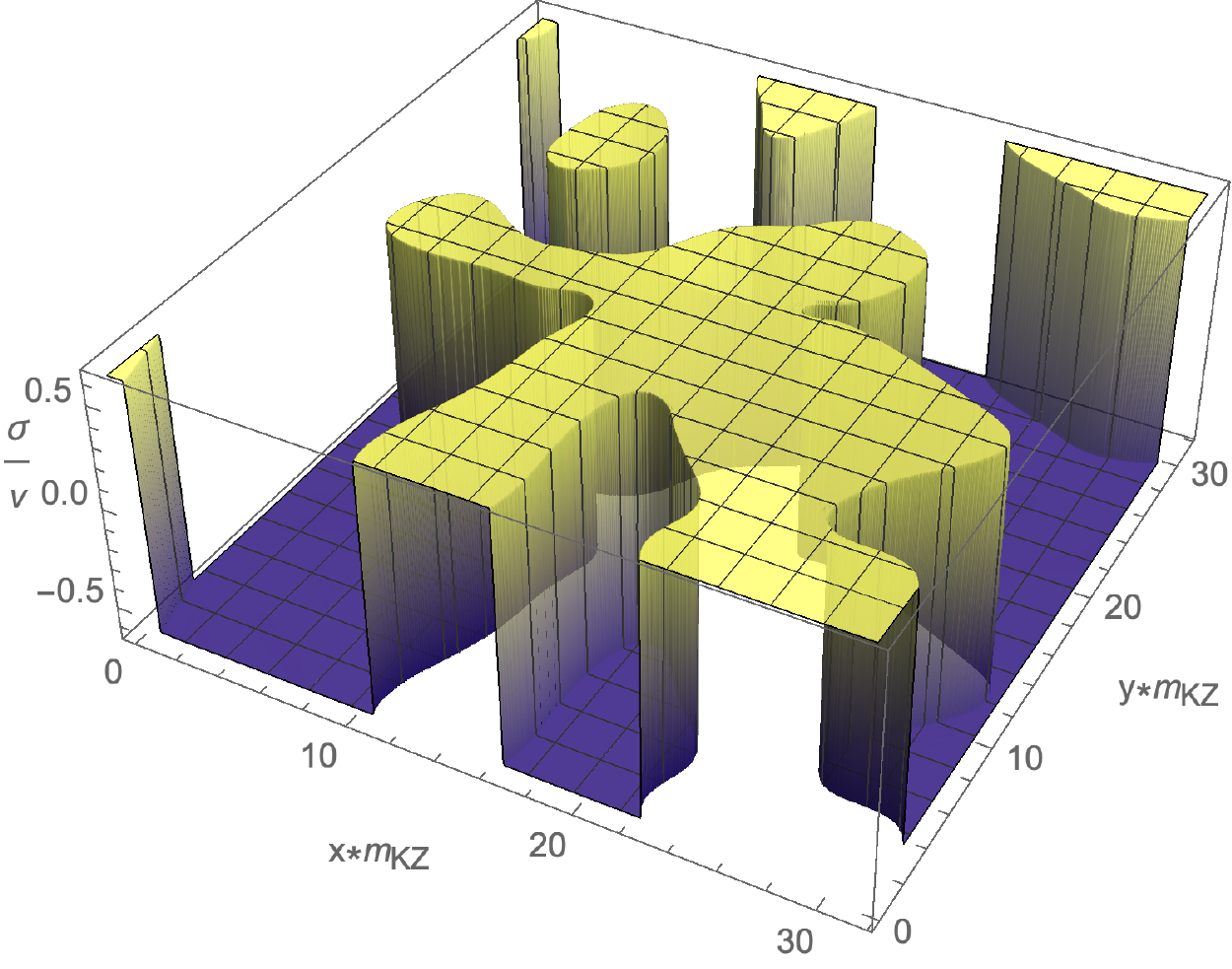}
	\caption{Evolution of the $\sigma$ field in scenario (B), the labels of the axes are the same in Fig.~\ref{fig:configA}. The parameters in the potential are the same as in Fig.~\ref{fig:sigma}. }\label{fig:configB}
\end{figure}

\bigskip

\section{GW prodction}


The Fourier modes of the tensor perturbation, $\sh_{ij}(\tau,\bk)$, satisfy the linearized Einstein equation
\bea
\sh''_{ij}(\tau, \bk) +  \frac{2a'}{a} \sh_{ij}'(\tau, \bk) + k^2 \sh_{ij}(\tau,\bk) = 16\pi G_N \tT_{ij}(\tau,\bk) \ , \nnn
\eea
where $\tT_{ij}$ is the transverse, traceless part of the energy-momentum tensor. 
During inflation, when $\sh(\tau,\bk)$ leaves the horizon $(k|\tau| < 1)$, they will be frozen to a fixed value $\sh^f(\bk)$. After inflation, when the mode reenters the horizon, $\sh^f$ will serve as the amplitude for later evolution. Then the general form of $\sh(\tau,\bk)$ is 
\bea
\sh_{ij} (\tau,\bk)= \sh^f_{ij}(\bk) \tilde{\cal E}^i_0(k) a^{-1} \sin(k\tau + \phi) \ ,
\eea
where $\tilde{\cal E}^i_0$ and $\phi$ are determined by the evolution of the universe before the mode reenters the horizon. The detailed form of $\tilde{\cal E}^i_0(k)$ and $\phi$ can be found in \cite{An:2022cce}. Then we calculate the GW energy density spectrum, 
\bea\label{eq:rhoGW}
\frac{d \rho_{\rm GW}}{d\ln k} = \frac{k^5 |\tilde{\cal E}^i_0(k)|^2}{64\pi^3 G_N a^4} \frac{\langle |\sh^f_{ij}(\bk)|^2 \rangle }{V} \ ,
\eea
where $G_N$ is the Newton gravity constant, and $V$ is the total comoving volume which will be canceled by the same volume factor in $\langle |\sh^f_{ij}(\bk)|^2 \rangle$.

\begin{figure}[htpb]
	\centering
	\includegraphics[width=0.8\linewidth]{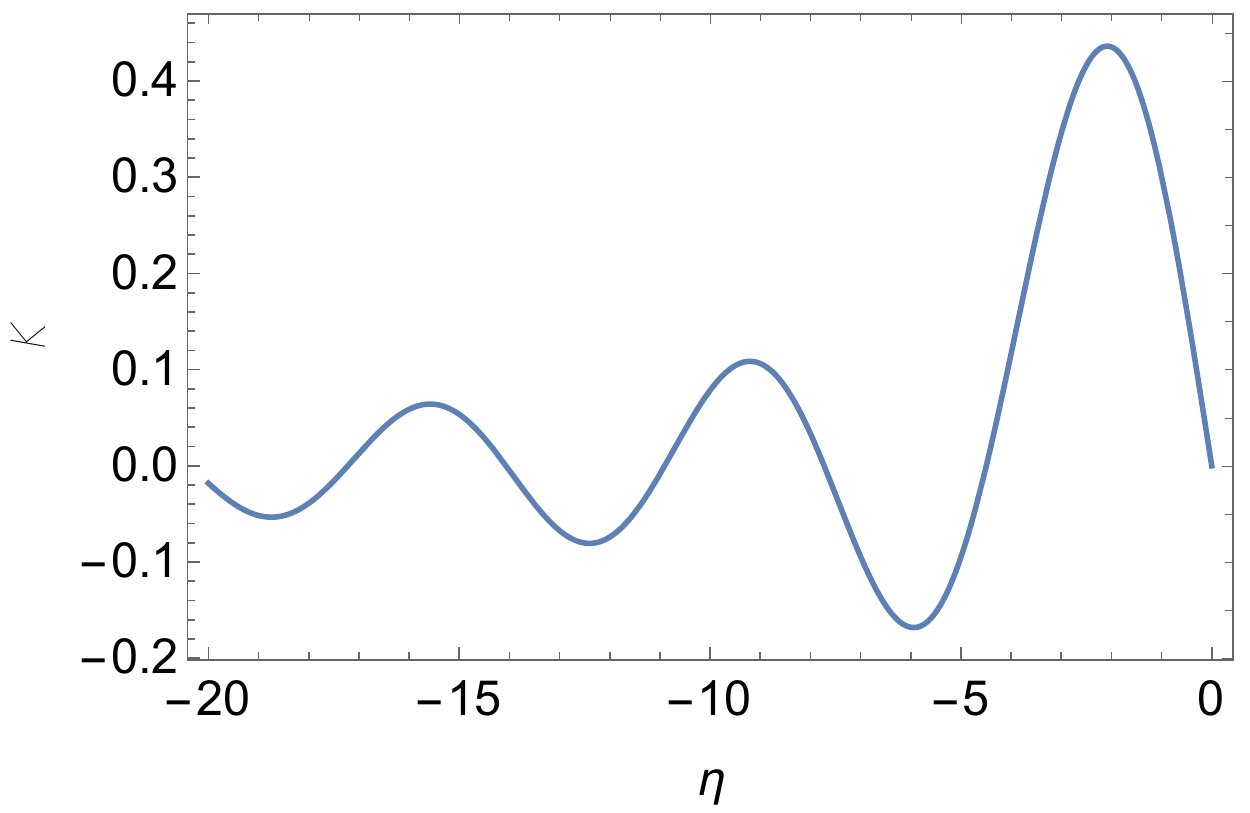}
	\caption{The kernel ${\cal K}$ as a function of $\eta \equiv k\tau$.}\label{fig:K}
\end{figure}

Using the Green's function method, we can calculate $\sh^f$~\cite{An:2020fff,An:2022cce},
\bea\label{eq:hf}
\sh^f_{ij} (\bk) = \frac{16\pi G_N}{k} \int_{-\infty}^0 d\tau' {\cal K}(k\tau') \tT_{ij}(\tau',\bk) \ ,
\eea
where ${\cal K}$ is the integral kernel, 
\bea\label{eq:K}
{\cal K}(\eta) = \frac{1}{\eta} \left( \cos\eta - \frac{\sin\eta}{\eta} \right) \ .
\eea
Eq.~(\ref{eq:hf}) indicates that during inflation because the time translation symmetry is badly broken, GWs can even be produced by a stationary source. The plot of ${\cal K}$ is shown in Fig.~\ref{fig:K}, from which we can see that the highest peak of ${\cal K}$ is around $k\tau' \approx -2$, and this is the moment $\tT_{ij}(\tau',\bk)$ contributes most to $\sh^f_{ij}(\bk)$.


\begin{figure}[htpb]
	\centering
	\includegraphics[width=1\linewidth]{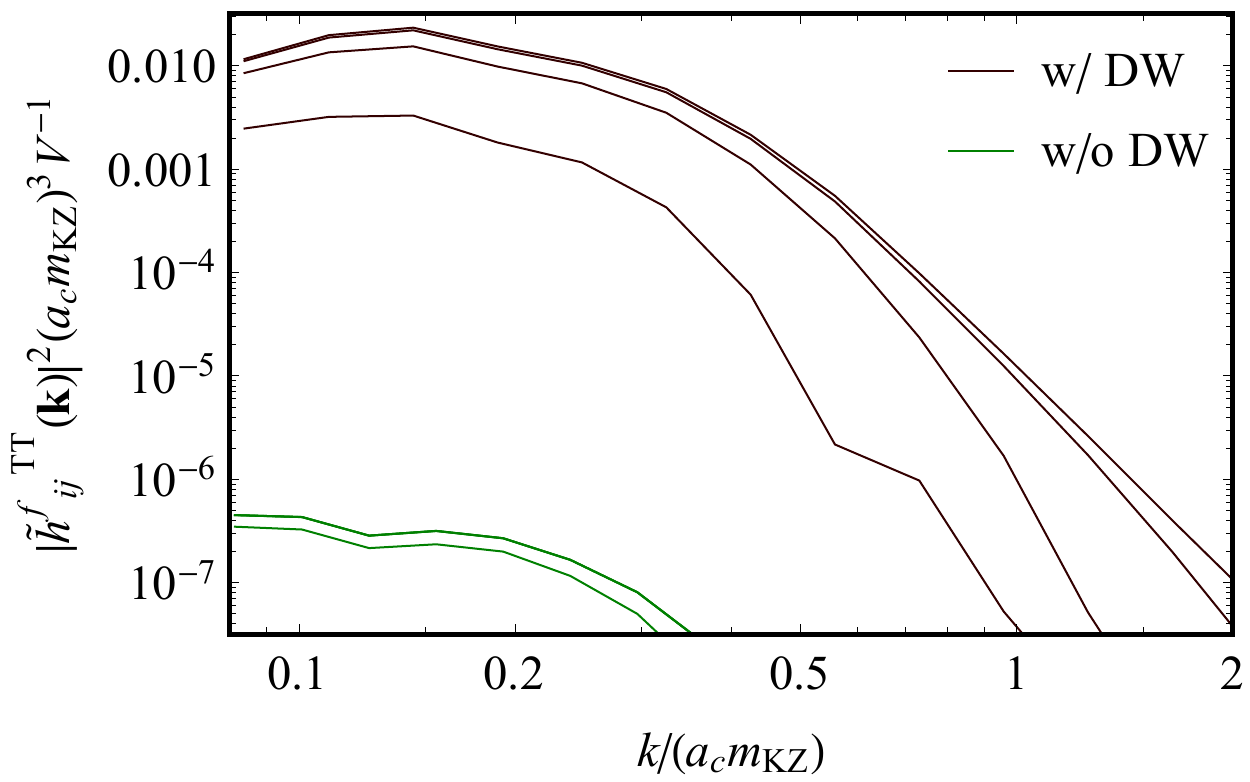}
	\includegraphics[width=1\linewidth]{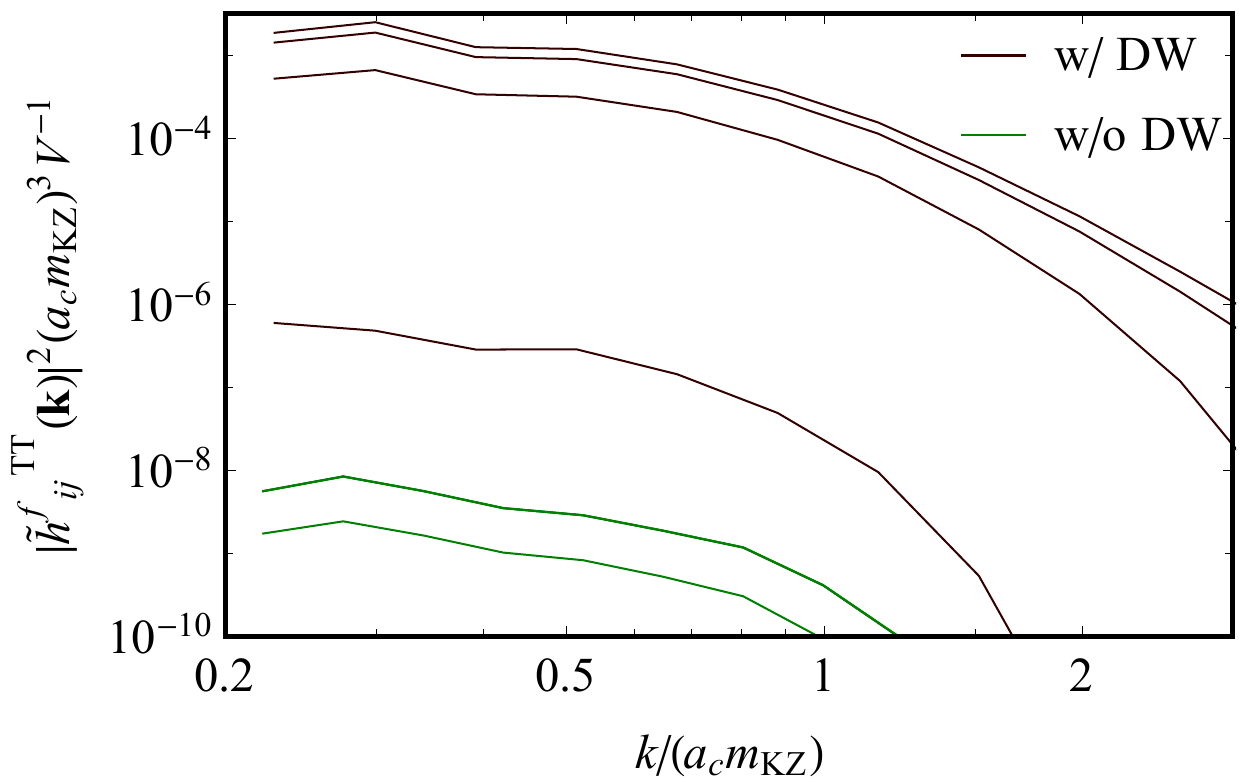}
	\caption{The spectrum of $\langle |\sh^f_{ij}|^2 \rangle$ for scenarios (A) (up) and (B) (down). 
The choices of parameters in the potential are the same as in Fig.~\ref{fig:sigma}. 
The spectrum is shown accumulated up to $\ln(a/a_c) = 1.3$, 2.3, 3.3 and 4.3. As a comparison, the spectrum without DW production is also shown by the green curves, where for both (A) and (B), the $\ln(a/a_c)=2.3,3.3,4.3$ curves are identical, which indicates that the GW production ceases completely after $\ln(a/a_c) = 2.3$. }\label{fig:hf}
\end{figure}

After doing a small $k\tau'$ expansion in the integral (\ref{eq:hf}), we can see that the integral is finite at $\tau' = 0$, and thus the integral is mostly contributed in the region $k\tau' \sim -2$. 
Fig.~\ref{fig:hf} shows the spectrum of $|\sh^f_{ij}|^2/V$ accumulated from $\tau_c$ to $\ln(a/a_c) = 1.3,2.3,3.3,4.3$ for both scenarios (A) and (B). One can see that the GW productions for both scenarios stop at about four e-folds after the phase transition. Th
As comparisons, we also did simulations without DW formations. We use the same parameters in the simulations without DW formation to generate the initial configuration $\sigma(\bx)$ induced by the exponential growth. Then we replace $\sigma(\bx)$ with $|\sigma(\bx)|$ for the initial condition for the nonlinear growth. This way, we eliminate all the DWs while keeping the initial fluctuations of the $\sigma$ field. The green curves in Fig.~\ref{fig:hf} show the accumulated $|\sh_{ij}|^2/V$ without DWs for both scenarios (A) and (B), in which the green curves for $\ln(a/a_c) = 2.3,3.3,4.3$ are identical in both scenarios. The coincidences indicate that without DWs, the GW production stops at about $\ln(a/a_c) = 2.3$ because the oscillations of $\sigma$ are quickly damped by the Hubble expansion, as shown in Fig.~\ref{fig:sigma}. This also agrees with the result in \cite{Garcia-Bellido:2007fiu}. From Eq.~(\ref{fig:hf}), one can also see that without DWs, the GW spectrum would be significantly smaller. Therefore, in the simulation, the DWs are the dominant source for the SGWB. 
 
\bigskip

\section{Detectability of the GW signal}

Today's GW relative abundance can be calculated from Eq.~(\ref{eq:rhoGW}), 
\bea
\Omega_{\rm GW}(f) = \Omega_R \times \rho_R^{-1} \left.\frac{d\rho_{\rm GW}}{d\ln f} \right|_{\rm today} \ ,
\eea
where $\Omega_R$ and $\rho_R$ are today's radiation abundance and energy density. $f = k/(2\pi a_{\rm today})$ gives today's GW frequency.
The detailed shape of $\Omega_{\rm GW}(f)$ also depends on the universe's evolution when the GW modes reenter the horizon. From the end of inflation to the completion of reheating, the universe may undergo some transition eras, such as matter domination and kination domination~\cite{Spokoiny:1993kt,Peebles:1998qn}. As shown in the appendix (see also in \cite{An:2022cce}), the peak value of $\Omega_{\rm GW}(f)$ is not sensitive to the universe's evolution when the modes are outside the horizon. During kination domination, the total energy density of the universe drops as $a^{-6}$, whereas the energy density of GW drops as $a^{-4}$, and therefore $\Omega_{\rm GW}(f)$ gets a relative enhancement~\cite{Co:2021lkc,Gouttenoire:2021jhk}. On the other hand, if the GW modes reenter the horizon in an intermediate matter dominated era before reheating, the total energy of the universe drops as $a^{-3}$. Thus $\Omega_{\rm GW}$, in this case, obtains a relative suppression. 

From qualitative arguments and numerical simulations in the appendix, we can derive a semi-analytical formula for the peak value of $\Omega_{\rm GW}$,
\bea\label{eq:analytical}
\Omega_{\rm GW}^{\rm peak} = \Omega_R \times {\cal A} (H d_w)^2 \times \left( \frac{\Delta\rho}{\rho_{\rm inf}} \right)^2 z_{\rm mp}^\alpha \ ,
\eea
where $d_w$ is the physical wall thickness at $\tau = - 2/\mkz$. $z_{\rm mp}$ in (\ref{eq:analytical}) measures the redshift of the universe between the time the peak mode reenters the horizon and the end of the intermediate stage. The following table presents the values of ${\cal A}$ and $\alpha$ for scenarios (A) and (B) and for different intermediate stages before reheating.

\begin{table}[htbp]
	\centering
	\label{tab}
	\begin{tabular}{cccc}
		\hline
		$\mathcal{A}$ ~&~ ~~~IRH ~&~ MD ($\alpha=-1$) ~&~KD ($\alpha=2$)\\
		\hline
		\hline
		Scenario A ~&~~~~~~ 0.15 ~&~ 0.09 ~&~ 0.2\\
		\hline
		Scenario B ~&~~~~~~ 0.3 ~&~ 0.15 ~&~ 0.3\\
		\hline
	\end{tabular}
\end{table}

\begin{figure}[htpb]
	\centering
	\includegraphics[width=0.96\linewidth]{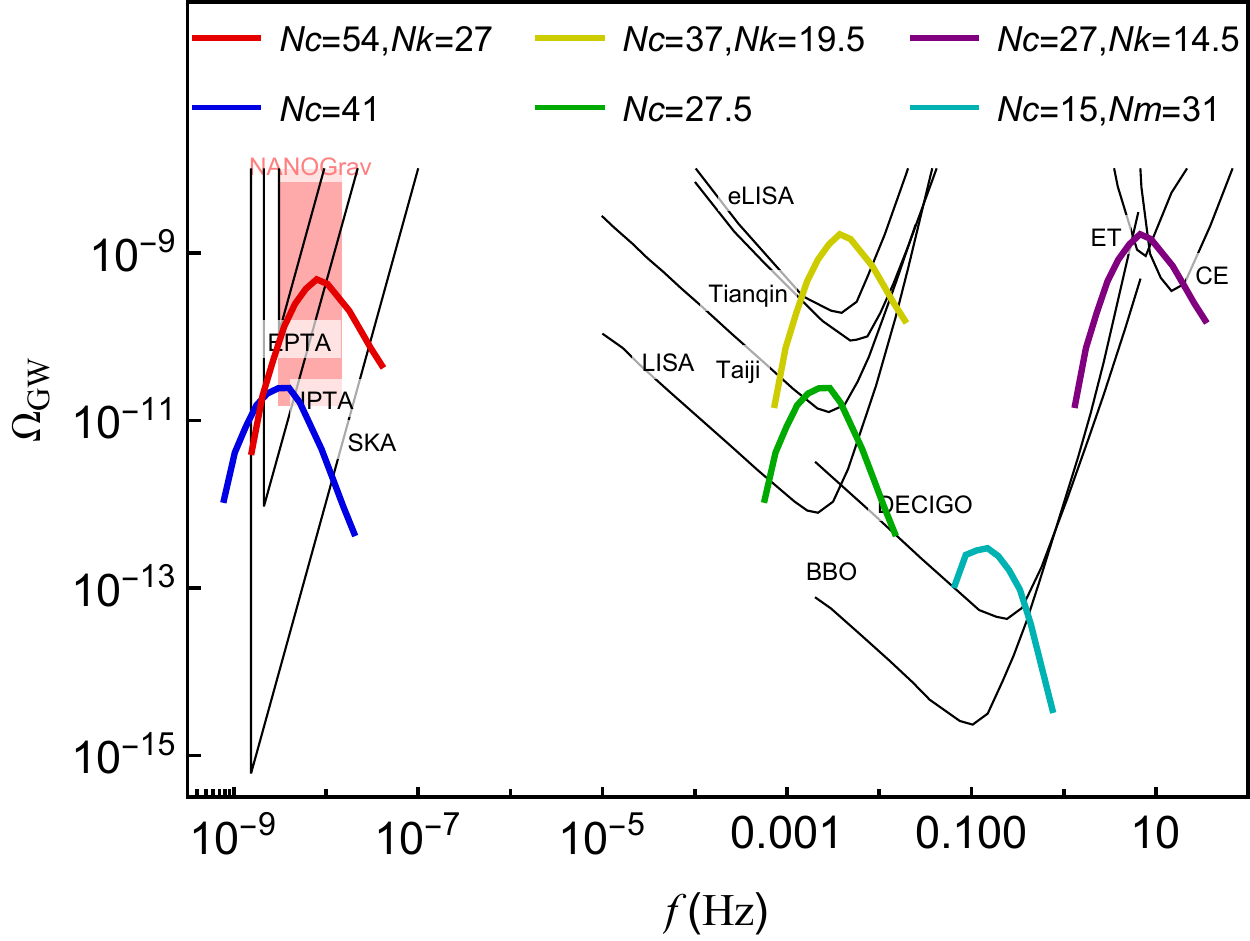}
	\caption{Spectrums of today's SGWB energy density distribution for scenario (B), together with the sensitivity curves of future GW detectors and the region favored by the NanoGrav observation. }\label{fig:signals}
\end{figure}

\begin{figure}[htpb]
	\centering
	\includegraphics[width=1\linewidth]{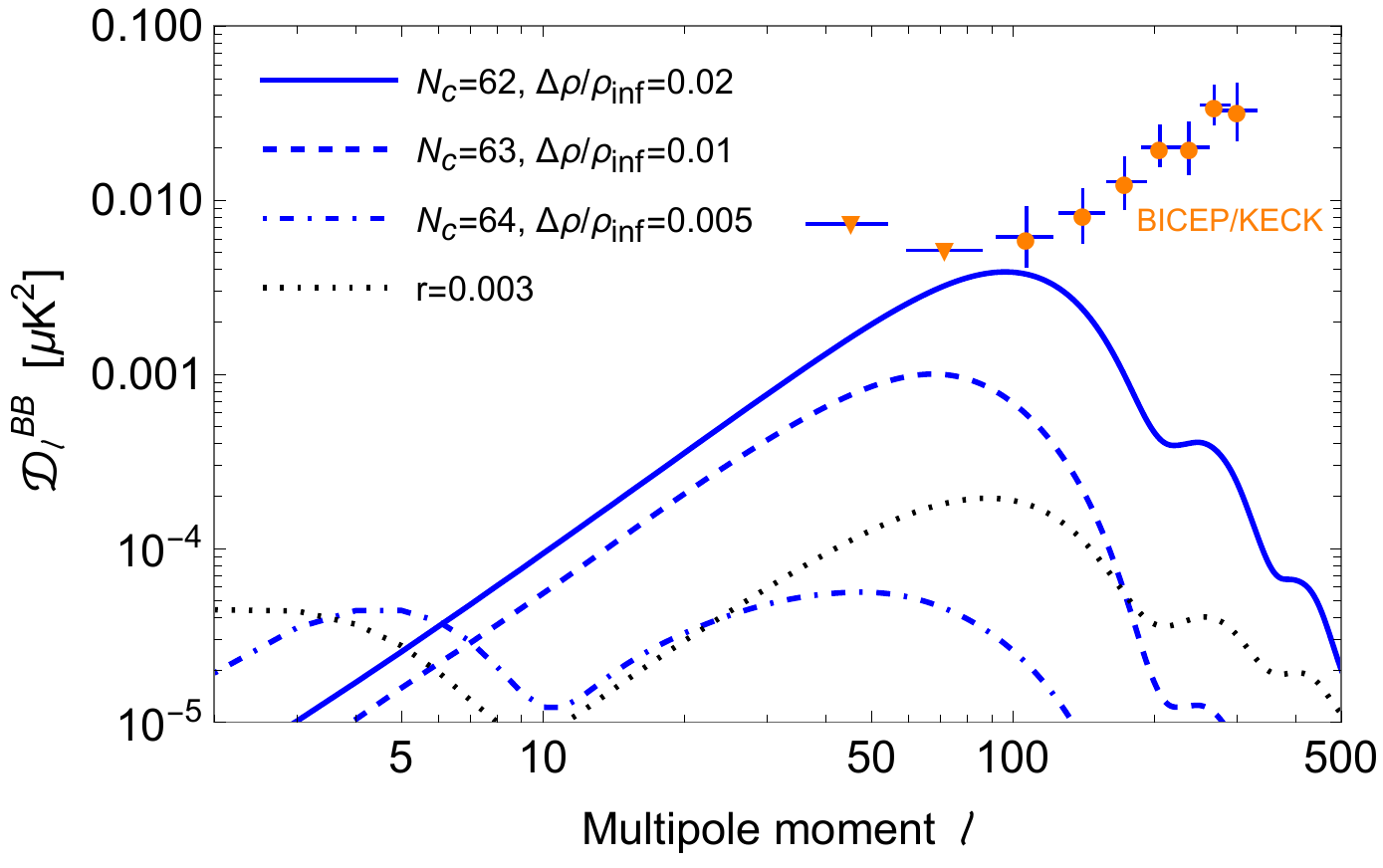}
	\caption{B-mode power spectrums induced by the SGWB in scenario (A) for different choices of phase transition time and potential energy in the spectator sector. As a comparison, the B-mode spectrum generated via tensor mode quantum fluctuations with tensor-scalar ratio $r=0.003$ is also shown. The orange dots and triangles are for the data from BICEP/KECK array. }\label{fig:Bmode}
\end{figure}

Future GW detectors can detect SGWB produced during inflation by the DWs. Fig.~\ref{fig:signals} shows today's SGWB spectrums and the sensitivities of PTAs and space and ground-based GW telescopes. Here we consider scenario (B), and the parameters for the spectator sector and inflaton model are the same as for the lower panel of Fig.~\ref{fig:hf}. We consider three after-inflation-evolution scenarios, the instantaneous reheating, intermediate MD, and intermediate KD. As shown in Fig.~\ref{fig:signals}, for $N_c \approx 40$, the SGWB spectrum falls in the region of PTAs, and it is possible to use it to explain the common noise process. For $N_c \approx 20-25$, the SGWB can be detected by the planned space GW detectors, such as LISA, Taiji, and Tianqin. For $N_c \approx 10$, it can be detected by future terrestrial GW detectors. 

If $N_c \approx 60$, the GW frequency can be about $10^{-19}$ Hz and can be detected via the CMB B-mode spectrum. Here we consider scenario (A), in which the back reaction to the curvature perturbation will not ruin the temperature correlation in the CMB spectrum~\cite{Jiang:2015qor}. With reasonable choices of parameters, the future CMB-S4 experiment can detect the signal.  Furthermore, we can also see that the shape of the B-mode spectrum induced by the phase transition differs from the quantum-induced one. Therefore, future CMB observations will be able to distinguish the origin of the SGWB source.

\section{Summary and discussions}

In this work, we numerically study the SGWB produced by DWs during inflation. We show that due to the severe violation of the time translation symmetry during inflation, even the cosmologically static DW configurations, without forming a network, can create detectable SGWB for future GW detectors. We also show the SGWB produced in scenarios can be used to explain the common red noise observed by NanoGrav. 

To emphasize the DW-induced SGWB during inflation, we assume that the DWs decay sufficiently fast such that the DW network disappears before dominating the universe, and the GWs produced by the DW network are negligible. 

Scenario (A) discussed in this work may also be realized in warm inflation models, where the decay of the inflaton field supports the thermal plasma~\cite{Berera:1995wh,Berera:1995ie}.
In scenario (B), the interaction between the $\phi$ and $\sigma$ leads to a back reaction to curvature perturbation, and the Lagrangian can be written as $y \sigma^2 \phi_0 \delta\phi$. This term may induce secondary GWs once the induced curvature perturbations reenter the horizon. This term may also lead to the production of primordial back holes. We leave these interesting phenomena to future studies.

\section{acknowledgments}

\begin{acknowledgments}
We thank Lian-Tao Wang and Zhong-Zhi Xianyu for helpful discussions. 
The work of HA is supported in part by the National Key R$\&$D Program of China under Grant No. 2021YFC2203100 and 2017YFA0402204, the NSFC under Grant No. 11975134, and the Tsinghua University Initiative Scientific Research Program.
\end{acknowledgments}

\bigskip

	\clearpage
	
	\setcounter{equation}{0}
	\setcounter{figure}{0}
	\setcounter{table}{0}
	\setcounter{section}{0}
	\setcounter{page}{1}
	\makeatletter
	\renewcommand{\theequation}{S\arabic{equation}}
	\renewcommand{\thefigure}{S\arabic{figure}}
	\renewcommand{\thetable}{S\arabic{table}}

\onecolumngrid
\begin{center}
	\textbf{\large Appendix}\\[0.3cm]
	\vspace{0.05in}
\end{center}

In the Appendix, we present the details of our numerical simulation of the formation of the DWs and calculation of the SGWB. Additionally, we derive a semi-analytical formula for the peak value of the SGWB spectrum. Thirdly, we present the detailed calculation of the influence on the SGWB spectrum induced by the evolution of the universe after inflation. 
 \\[0.3cm]
	
\twocolumngrid	

\section{Details of the numerical simulation}

Here we present the detailed simulation of the evolution of the spectator field $\sigma$ during the second-order phase transition. 

We use comoving coordinate with the conformal time, and the background de Sitter metric can be written as
\bea
ds^2 = a^2(\tau) (d\tau^2 - dx^2 - dy^2 - dz^2) \ .
\eea
where $a(\tau) = - 1/ H\tau$ is the scale factor. The Lagrangian of the spectator $\sigma$ can be written as
\bea
{\cal L} = \frac{1}{2H^2 \tau^2} \left[ \sigma'^2 - \nabla\sigma\cdot\nabla\sigma \right] - \frac{1}{H^4\tau^4} V(\sigma) \ ,
\eea 
where, the primes indicate derivatives with respect to $\tau$.
In this work, we assume the potential is in the Landau-Ginzburg form, 
\bea
V = - \frac{1}{2} m_{\rm eff}^2 \sigma^2 + \frac{\lambda}{4} \sigma^4  \ . 
\eea
We discuss two scenarios. In scenario (A), the phase transition is triggered by evolution of the temperature at the beginning of inflation, and $m_{\rm eff}^2 = y T^2 - m^2$. In scenario (B), the phase transition is triggered by the evolution of the inflaton field, and $m_{\rm eff}^2 = y \phi^2 - m^2$. Around the critical time $\tau_c$, the potential can be written as 
\bea
V_{\rm KZ} = - \frac{1}{2} \mkz^3 a_c^{-1} (\tau - \tau_c) \sigma^2 + \frac{\lambda}{4} \sigma^4 \ .
\eea

\subsection{Tachyonic growth of the long wavelength modes}


Now we consider the quantum behavior of $\sigma$ at the beginning of the phase transition. Around $\tau_c$, the $\sigma$ field still locates at $\sigma = 0$, and thus the interaction term can be neglected. As a result, $m_{\rm eff}^2 < 0$, and the long wavelength $\sigma$ modes grow tachyonically. Here we generalize the discussions in \cite{Garcia-Bellido:2002fsq} to dS space. 

The Hamiltonian governs the evolution of $\sigma$ in the linear region is 
\begin{equation}
	\begin{split}
		\mathscr{H}&=\int d^3\bx \frac{1}{2}\left[\frac{1}{a^2}\pi^2+a^2(\nabla \sigma)^2+a^4 m_{\text{eff}}^2 \sigma^2\right]\\
		&=\int \frac{d^3 \bk}{(2\pi)^2} \frac{1}{2}\left[\frac{1}{a^2}\tilde{\pi}^2+\left(a^2 k^2+a^4 m_{\text{eff}}^2 \right)\tilde{\sigma}^2\right] \ ,
	\end{split}
\end{equation}
where
\begin{equation}
	\pi\equiv a^2 \phi' 
\end{equation}
is the canonical momentum.
Then the Fourier modes of $\sigma$ and $\phi$ satisfy the Hamiltonian equations of motion
\begin{equation}
	\frac{d}{d \tau}\left(\begin{matrix}
		\tilde{\pi}\\
		\tilde{\sigma}
	\end{matrix}\right)=\left(\begin{matrix}
		0&-\left(a^2 k^2+a^4 m_{\text{eff}}^2 \right)\\
		a^{-2}&0
	\end{matrix}\right)\left(\begin{matrix}
		\tilde{\pi}\\
		\tilde{\sigma}
	\end{matrix}\right) \ .
\end{equation}
To quantize the theory, we impose the equal-time  canonical commutation relation to $\sigma$
\begin{equation}
	\left[\sigma(\bx,\tau),\pi(\mathbf{y},\tau)\right]=i\delta^3(\bx-\mathbf{y}) \ ,
\end{equation}
We expand $\sigma$ by ladder operators $a_\bk,\ a^\dagger_{-\bk}$,
The solution can be expressed as
\begin{align}
	\tilde{\pi}(\bk,\tau)&=a_\bk a(\tau)^2f'(k,\tau)+a^\dagger_{-\bk}a(\tau)^2f'^*(k,\tau),\\
	\tilde{\sigma}(\bk,\tau)&=a_\bk f(k,\tau)+a^\dagger_{-\bk}f^*(k,\tau).
\end{align}
where the mode function $f$ depends only on $k$ since the space is homogeneous and isotropic and satisfies the equation of motion
\bea\label{eq:ff}
f'' - \frac{2f'}{\tau} + \left(k^2 +\frac{m^2_{\rm eff}}{H^2 \tau^2} \right)f = 0 \ .
\eea

We require the ladder operators to satisfy the commutation relation
\begin{equation}
	\left[a_\bk,a^\dagger_{\bq} \right]=(2\pi)^3\delta^3(\bk-\bq)\ .
\end{equation}
This leads to the normalization condition for the mode function
\begin{equation}
	a(\tau)^2\left(f(k,\tau)f^{*\prime}(k,\tau)-f'(k,\tau)f^*(k,\tau)\right)=i \ .
\end{equation}

The time scale of the phase transition is determined by $\mkz$, which is much larger than $H$, and thus when setting the initial condition of $f$ at $\tau_c$, we can neglect the Hubble expansion. Therefore just as in \cite{Garcia-Bellido:2002fsq}, we choose
\begin{equation}\label{IC}
	f(k,\tau_c)=\frac{1}{a(\tau_c)}\frac{1}{\sqrt{2k}}e^{-ik\tau_c},\ \ \ \ f'(k,\tau_c)=\frac{-i}{a(\tau_c)}\sqrt{\frac{2}{k}}e^{-ik\tau_c} \ .
\end{equation}
as the initial condition for (\ref{eq:ff}). 

\begin{figure}[htpb]
	\centering
	\includegraphics[width=0.95\linewidth]{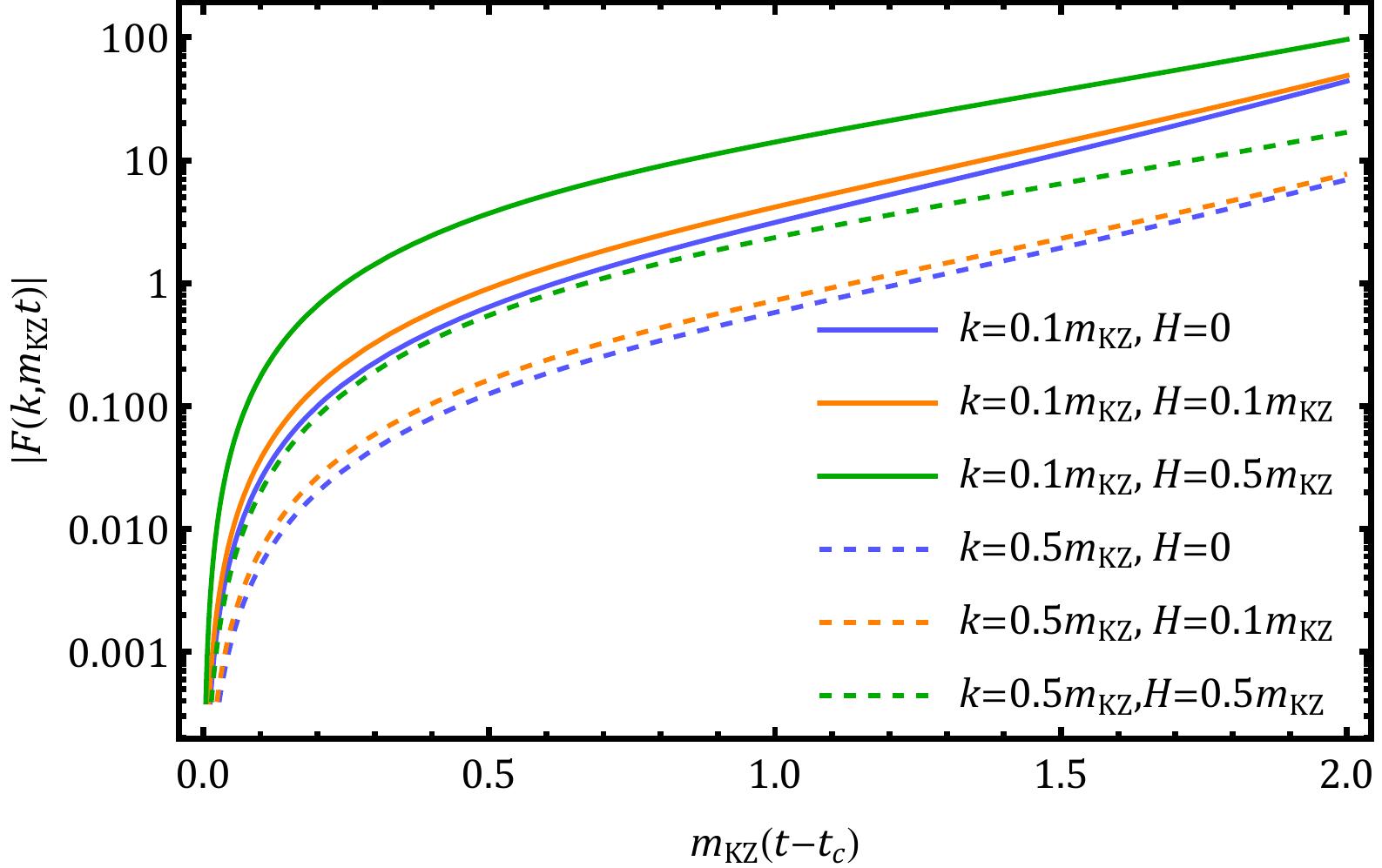}
	\caption{$F(k)$ as functions of physical time $t$ for different values of $H$. }\label{fig:Ft}
\end{figure}

The modes become classical when the anti-commutation of $\ts_\bk$ and $\tilde\pi_\bk$ is significantly larger than their commutation,
\bea
|\langle [\hat\ts_\bk(\tau), \hat{\tilde\pi}_\bk(\tau)]_+ \rangle | \gg |\langle [\hat\ts_\bk(\tau), \hat{\tilde\pi}_\bk(\tau)] \rangle | \ ,
\eea
where the expectation value is over the temporary vacuum state at $\tau_c$. This is equivalent to require 
\bea
F(k,\tau) \gg 1 \ ,
\eea
where the function $F$ is defined as 
\begin{equation}
	F(k,\tau)=a(\tau)^2{\rm Re}\left[f'(k,\tau)f^*(k,\tau)\right] \ .
\end{equation}
A plot of $F(k)$ is shown in Fig.~\ref{fig:Ft} for different values of $k$ and $H$. In this plot, to compare with Minkowski space ($H=0$) result in \cite{Garcia-Bellido:2002fsq}, we use the physical time $t$ in Fig.~\ref{fig:Ft}, defined as $dt = a d\tau$. One can see that $F$ grows exponentially after the phase transition for the modes with $k < \mkz$ as long as $H$ is significantly smaller than $\mkz$.

Then the canonical phase space distribution of the long-wavelength modes can be described by the Gaussian ensemble given by the Wigner function~\cite{Polarski:1995jg,Lesgourgues:1996jc,Kiefer:1998qe} 
\bea\label{eq:Wigner}
W_\bk (q_\bk,p_\bk)&=& \frac{1}{\pi^2}\exp \left[-\frac{|\sigma_\bk|^2}{|f(\bk,\tau)|^2} \right. \nnn
&&\!\!\!\!\left.-4|f(\bk,\tau)|^2\left|\pi_\bk-\frac{F(\bk,\tau)}{|f(\bk,\tau)|^2}\sigma_\bk\right|^2\right] \ .
\eea

\begin{figure}[htpb]
	\centering
	\includegraphics[width=0.9\linewidth]{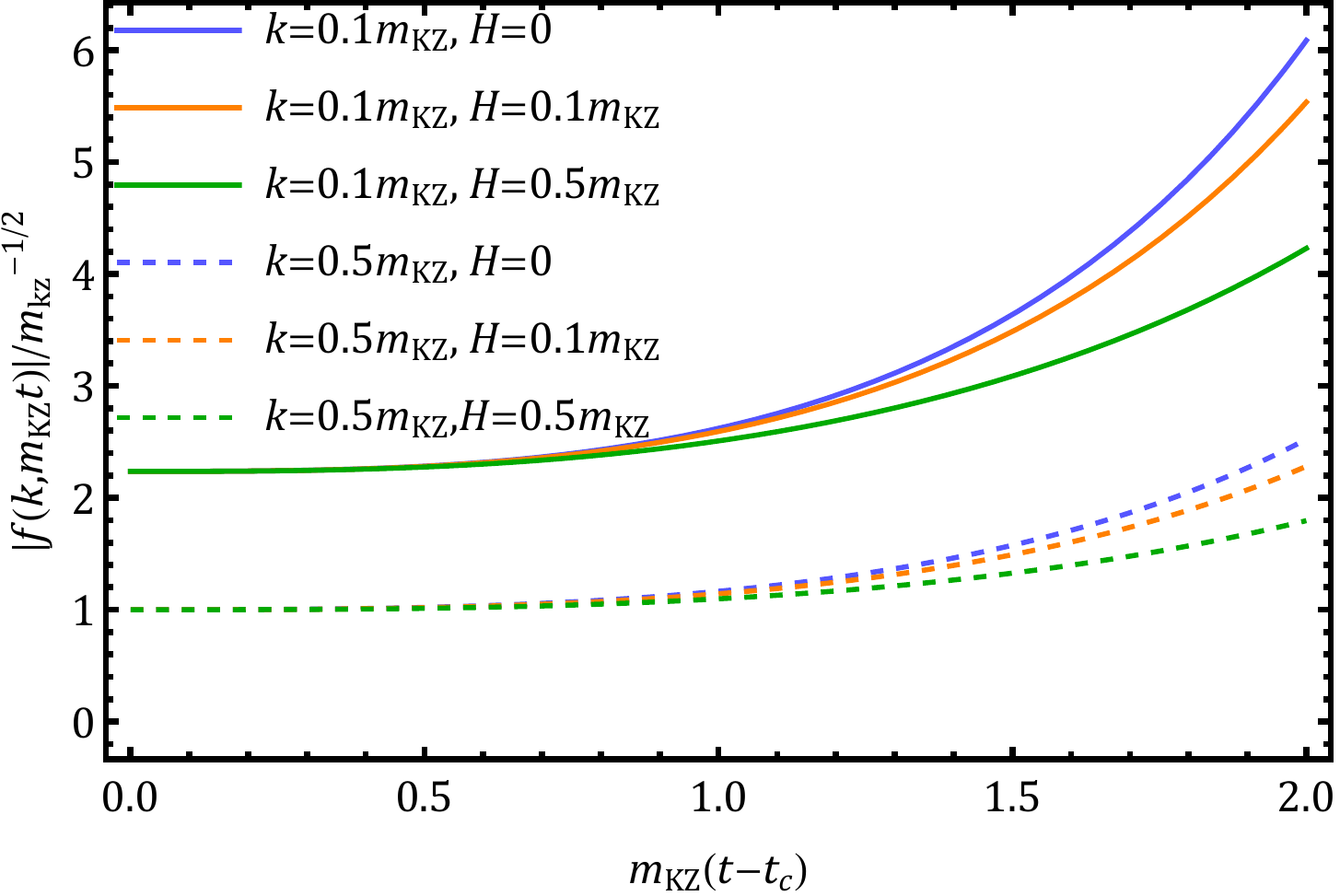}
	\caption{Evolution of the mode function $f(k)$ as a functions of the physical time $t$ for different values of $k$ and $H$. }\label{fig:fkt}
\end{figure}

\begin{figure}[htpb]
	\centering
	\includegraphics[width=0.9\linewidth]{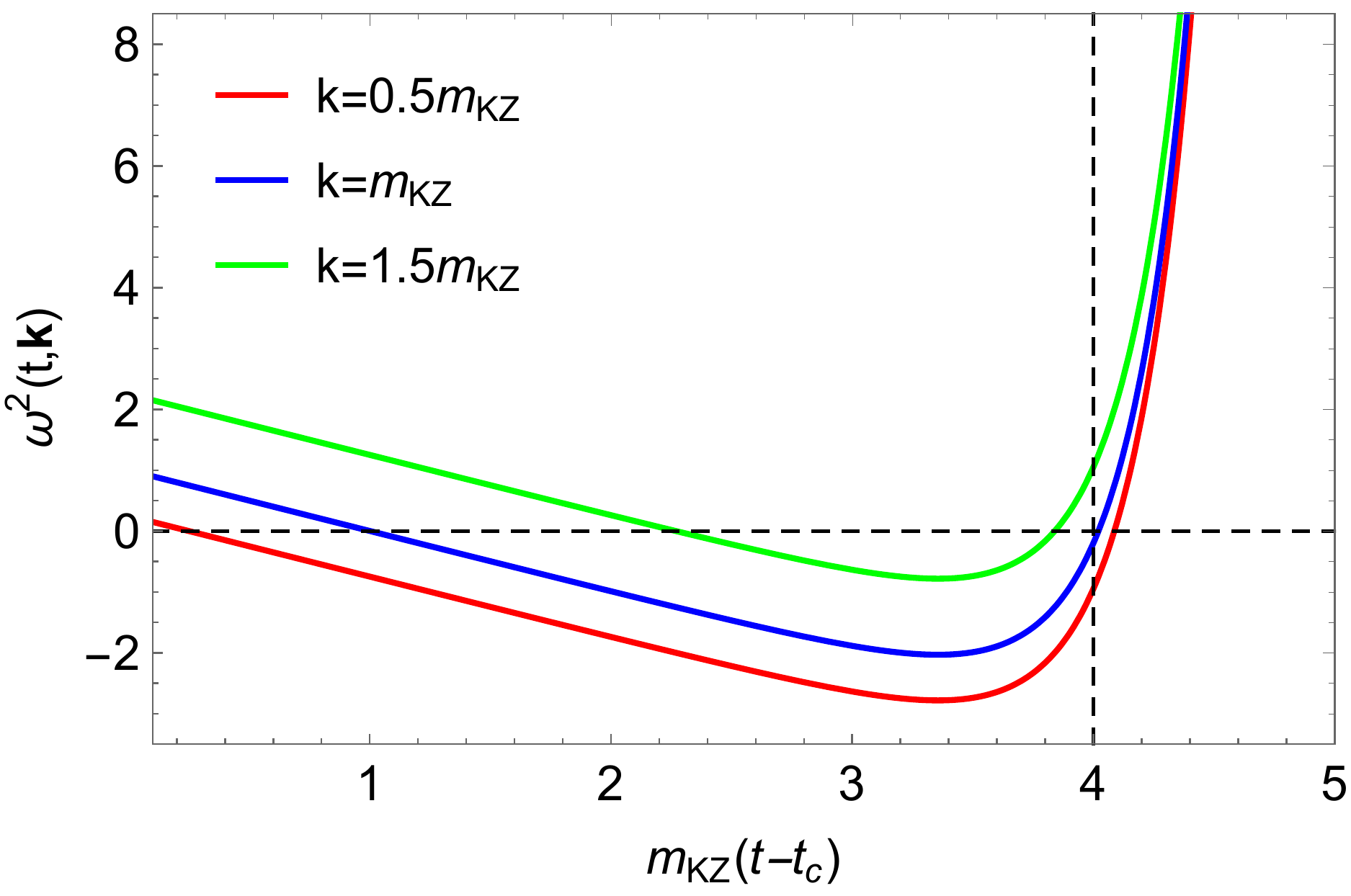}
	\caption{Evolution of the mode function $f(k)$ as a functions of the physical time $t$ for different values of $k$ and $H$. }\label{fig:omega2}
\end{figure}

After $\tau_c$, due to the tachyonic growth, the nonlinear term $
\lambda \sigma^4$ becomes more and more important, and then when we consider the evolution of $\ts(\bk)$, we must include the back reaction from the long-wavelength modes. Thus the linearized equation of motion for $\ts(\bk)$ can be written as
\bea\label{eq:sigmak}
\sigma_\bk'' + \frac{2 a'}{a} \sigma'_\bk + \omega^2_\bk(\tau) \sigma_\bk = 0 \ ,
\eea
with 
\bea\label{eq:omegak}
\omega_\bk^2 = k^2 - a_c^2 m_{\rm KZ}^3(\tau - \tau_c) + \frac{\lambda}{2} \langle \sigma^2(\tau,\bx)\rangle \ ,
\eea
where the expectation value of $\langle\sigma^2(\tau,\bx)\rangle$ includes all the contributions from the long-wavelength modes that have grown exponentially. Fig.~\ref{fig:omega2} shows the evolution of $\omega^2$ for different values of $k$ with $H = 0.1\mkz$ and $\lambda = 0.1$.
Note that $\sigma(\bk)$ can grow tachyonically only when $\omega^2 < 0$. Thus one can see that only modes with $k < {\cal O}(\mkz)$ can grow tachyonically. This determines that the size of the topological structure is about $2\pi/\mkz$ as shown Figs.~\ref{fig:configA} and \ref{fig:configB}. 

We should note that the time span for $\omega^2 < 0$ is not sensitive to the value of $\lambda$, since after the phase transition, $\langle\sigma^2(\bx) \rangle$ grows as $e^{[\mkz(t - t_c)]^3}$. 

%

Follow \cite{Garcia-Bellido:2002fsq}, we choose to use $\mkz(t-t_c) = 2$ as the matching point between tachyonic growth and the nonlinear classical evolution. In detail, we solve numerically Eq.~(\ref{eq:ff}) with the initial condition (\ref{IC}). Then at $\mkz(t - t_c) = 2$, we randomly generate $\ts(\bk)$ and $\tilde\pi(\bk)$ according to the distribution function (\ref{eq:Wigner}). Then from $\ts$ and $\tilde\pi$ we calculate the initial configuration of $\sigma(\bx)$ and $\pi(\bx)$ for the classical lattice simulation of the nonlinear evolution.

\subsection{Nonlinear classical evolution}

The evolution of classical systems can be described by the Hamiltonian equations, 
\begin{align}
	\sigma'&=\frac{\delta \mathscr{H}}{\delta \pi}=\{\sigma,\mathscr{H}\},\\
	\pi'&=-\frac{\delta \mathscr{H}}{\delta \sigma}=\{\pi,\mathscr{H}\},
\end{align}
with the Poisson bracket $\{\ ,\ \}\equiv \int d^3x \frac{\delta}{\delta \sigma}\frac{\delta}{\delta \pi}-\frac{\delta}{\delta \pi}\frac{\delta}{\delta \sigma}$. Define $D_\mathscr{H}=\{\cdot,\mathscr{H}\}$, the solutions of Hamiltonian equations are
\begin{eqnarray}
	\sigma(\tau)&=e^{\int D_\mathscr{H} d \tau}\sigma(\tau_0),\\
	\pi(\tau)&=e^{\int D_\mathscr{H} d \tau}\pi(\tau_0).
\end{eqnarray}
It's convenient to use the dimensionless variables
\begin{equation}
	\Sigma=\frac{\sigma}{v},\ \Pi=\frac{1}{m}\frac{\pi}{v},\ M_{\text{eff}}=\frac{m_{\text{eff}}}{m} ,\ X=mx,\ T=m\tau \ .
\end{equation}
Then we have
\begin{eqnarray}
	\Sigma(T)&=e^{\int D_{\mathscr{H}/m} d T}\Sigma(T_0),\\
	\Pi(T)&=e^{\int D_{\mathscr{H}/m} d T}\Pi(T_0).
\end{eqnarray}
For the case studied in this work, the Hamiltonian can be divided into two parts,
\begin{equation}
	\mathscr{H}=\mathscr{K}+\mathscr{V},
\end{equation}
with
\begin{align}
	\mathscr{K}&=\frac{m}{\lambda}\int d^3 \CX \frac{1}{2}a^{-2}\Pi^2,\\
	\mathscr{V}&=\frac{m}{\lambda}\int d^3 \CX \frac{1}{2}a^{2}\left(\frac{\partial \Sigma}{\partial \CX}\right)^2+a^4 \left(\frac{1}{2}M_{\text{eff}}^2\Sigma^2+\frac{1}{4} \Sigma^4\right).
\end{align}
We use the symplectic integrator method~\cite{Forest:1989ez,Yoshida:1990zz} to numerically calculate the evolution of $\Sigma$. Define
\begin{align}
	A&=\Delta T^{-1} \int_{T}^{T+\Delta T} D_{\mathscr{K}/m} d T,\\
	B&=\Delta T^{-1} \int_{T}^{T+\Delta T} D_{\mathscr{V}/m} d T,
\end{align}
Here the infinitesimal evolutions $A$ and $B$ are generated by the dimensionless kinetic energy potential energies $\mathscr{K}/m$ and $\mathscr{V}/m$. 
An $n$th order symplectic integrator is defined as a symplectic approximation of the evolution operator with a difference of order $\mathcal{O}(\Delta T^{n+1})$ 
\begin{equation}
	e^{\int D_{\mathscr{H}/m} d T}=e^{(A+B)\Delta T}=\prod_{i=1}^{k}e^{c_i A\Delta T}e^{d_i B\Delta T}+\mathcal{O}(\Delta T^{n+1}) , 
\end{equation}
where $k = 8$ for sixth order accuracy used in this work. 
In our simulation we use the Solution A of 6th order symplectic integrator proposed in ~\cite{Yoshida:1990zz}, and the values for the coefficients are
\begin{align}
	d_1&=d_7=0.784513610477560,\\ d_2&=d_6=0.235573213359357,\\ d_3&=d_5=-1.17767998417887,\\ 
	d_4&=1-2(d_1+d_2+d_3),\ d_8=0,\\
	c_1&=c_8=\frac{d_1}{2},\ c_2=c_7=\frac{d_1+d_2}{2},\\ c_3&=c_6=\frac{d_2+d_3}{2},\ c_4=c_5=\frac{d_3+d_4}{2}.
\end{align} 

On lattice, derivative is replaced by difference. We use the neutral differential defined as
\bea
\Delta_xf(n)&=&\frac{8[f(n+1)-f(n-1)]-[f(n+2)-f(n-2)]}{12\delta x}\nn \ .
\eea
The Laplacian is defined as
\begin{equation}
	\begin{split}
		&\Delta_x^2 f(n)\\
		=&\frac{16[f(n+1)\!+\!f(n-1)]\!-\![f(n+2)\!+\!f(n-2)]\!-\!30f(n)}{12 \delta x^2}\\
	\end{split}
\end{equation}
Both $\Delta_x$ and $\Delta_x^2$ are at ${\cal O}(\delta x^4)$ accuracy.
$\Delta_x f(n)$ can be written as a general form
\begin{equation} 
	\Delta_{x}f(n)=\frac{1}{\delta x}\sum_{m}D(n-m)f(m) \ ,
\end{equation}
and in our case we have 
\begin{equation}\label{eq:Dn}
	D(n)=\frac{8(\delta_{n,-1}-\delta_{n,1})-(\delta_{n,-2}-\delta_{n,2})}{12} .
\end{equation}
With $D(n)$ we can define the effective momentum 
\begin{equation}
	k^{\rm eff}_{\tilde{n}}=\I \sum_{m}\frac{1}{\delta x}D(m) e^{-\I \frac{2\pi}{N}\tilde{n}m}.
\end{equation}
Then we define the transverse projective operator as~\cite{Huang:2011gf,Figueroa:2011ye}
\begin{equation}
	P_{ij}(\tilde{\mathbf{n}})=\delta_{ij}-k^{\rm eff}_i(\tilde{\mathbf{n}})k^{\rm eff}_j(\tilde{\mathbf{n}})/|\bk^{\rm eff}(\tilde{\mathbf{n}})|^2 \ 
\end{equation}
to suppress the unphysical modes in lattice calculation of the GWs. Specifically, in our choice of $D(n)$ in (\ref{eq:Dn}) we have
\begin{equation}
	k^{\rm eff}_{\tilde{n}}=\frac{1}{\delta x}\left(\frac{4}{3}\sin \left[\frac{2\pi}{N}\tilde{n}\right]-\frac{1}{6}\sin \left[\frac{4\pi}{N}\tilde{n}\right]\right).
\end{equation}

\section{Semi-analytical results}

Here we present a derivation to Eq.~(\ref{eq:analytical}) in the main text and discuss the IR and UV behavior of the SGWB spectrum, and compare the analytical result with numerical simulations.

Due to the shape of the kernel ${\cal K}$, as shown in Fig.~\ref{fig:K}, the dominant contribution of $\tT_{ij}(\tau',\bk)$ to $\sh^f_{ij}(\bk)$ happens at $k\tau' \approx -2$. To estimate the peak value of $\Omega_{\rm GW}$ we only keep this dominant contribution. Thus we have
\bea\label{eq:hhff}
\frac{\langle |\sh^f_{ij}|^2\rangle }{V} = {\cal A}_2 \left[\frac{16\pi G_N}{k^2}\right]^2 \frac{\langle |\tT_{ij}(-2/k,\bk)|^2\rangle}{V} \ .
\eea
Here $\langle\cdots\rangle$ means taking average over the direction of $\bk$. 

The space average value of ${T^{TT}}(\tau,x)^2$ can be written as
\bea\label{eq:TTT}
\frac{1}{V} \int d^3 x \tT_{ij}(\tau,\bx) \tT_{ij}(\tau,\bx) = \frac{1}{V} \int \frac{d^3k}{(2\pi)^3}  |\tT_{ij}(\tau,\bk)|^2 \ . \nnn
\eea
As discussed in the main text, the largest correlation of $\sigma$ field after the phase transition is determined by $\mkz$, and the typical size of independent regions in the universe is about $2\pi \mkz^{-1}$. Thus, to calculate the spatial integral in (\ref{eq:TTT}), we can first define a window function $W^{(a)}$ for each region. The center position of each region locates at $\bx^{(a)}$. Then the energy-momentum tensor can be decomposed as
\bea
T^{TT}_{ij}(\tau,\bx) = \sum_a \left(T^{TT}_{ij}(\tau,\bx)W^{(a)}(\bx)\right) \ .
\eea
Then we can define $\tilde T_{ij}^{TT(a)}(\tau,\bk)$ as
\bea
\tilde T_{ij}^{TT(a)}(\tau,\bk) = \int d^3xT^{TT}_{ij}(\tau,\bx)W^{(a)}(\bx) e^{-i \bk\cdot (\bx-\bx^{(a)})} \ . \nnn
\eea
Therefore, we have
\bea
\tT_{ij}(\tau,\bk) = \sum_a \tilde T^{TT(a)}_{ij}(\tau,\bk) e^{i \bk\cdot \bx^{(a)}} \ .
\eea
Then the righthand side of (\ref{eq:TTT}) can be written as
\bea
\frac{1}{V} \int\frac{d^3k}{(2\pi)^3} \sum_{a,b} \tilde T^{TT(a)}_{ij}(\tau,\bk) T^{TT(b)}_{ij}(\tau,\bk)^* e^{i \bk \cdot (\bx^{(a)} - \bx^{(b)})} \ . \nnn
\eea
The interference contributed from different patches is suppressed. Therefore it can be further simplified as 
\bea
\frac{1}{V} \int\frac{4\pi k^3 d\ln k}{(2\pi)^3} \left\langle\sum_{a} |\tilde T^{TT(a)}_{ij}(\tau,\bk) |^2\right\rangle \ . \nnn
\eea
In comoving coordinates the size of each independent region is about $2\pi \mkz^{-1} a_c^{-1}$. Therefore the above expression can be simplified as 
\bea\label{eq:A3}
{\cal A}_3 \left(\frac{\mkz a_c}{2\pi}\right)^3 \int \frac{4\pi k^3 d\ln k}{(2\pi)^3} |\tilde T^{TT(A)}_{ij}(\tau, k)|^2  \ ,
\eea
where ${\cal A}_3$ is a numerical factor and $|\tilde T^{TT(A)}_{ij}(
\tau, k)|^2$ can be seen as the space and angular averaged value of $|\tilde T^{TT(a)}_{ij}(\tau,\bk) |^2$. 

Now, let's consider the DW contribution to $\tT_{ij}$ in one region. $\tT_{ij}(\tau,\bk)$ can be expressed as 
\bea\label{eq:single}
\tT_{ij}(\tau,\bk) = \int\frac{d^3 p }{(2\pi)^3} \Lambda_{ij,kl} p^k p^l \ts(\bp) \ts(\bk - \bp) \ .
\eea
We know that the modes with $k \sim \mkz a_c$ dominates the contribution to the SGWB and as we will see later the $p$ integral in (\ref{eq:single}) is dominated by the region $p\sim m \gg \mkz$. Therefore, as a qualitative estimate, we can neglect the $\bk$ in $\ts$. Then the angular average of $\tT_{ij}(\tau,\bk)\tT_{ij}(\tau,\bk)$ contributed by a single region can be written as
\bea
\int\!\!\!\frac{d^3p}{(2\pi)^3}\!\!\int\!\!\!\frac{d^3q}{(2\pi)^3}\!\!\left[\int \frac{d\hat\bk}{4\pi} \Lambda_{ij,kl}\Lambda_{ij,k'l'} p^k p^l q^{k'} q^{l'}\right] \!\!|\ts(\bp)|^2 |\ts(\bq)|^2 \nnn.
\eea
Simple calculation gives
\bea\label{eq:ppqq}
\int \frac{d\hat\bk}{4\pi} \Lambda_{ij,kl}\Lambda_{ij,k'l'} p^k p^l q^{k'} q^{l'} = \frac{2}{5}(\bp\cdot \bq)^2 - \frac{2}{15}p^2 q^2 \nn
\eea

\begin{figure}[htpb]
	\centering
	\includegraphics[width=0.9\linewidth]{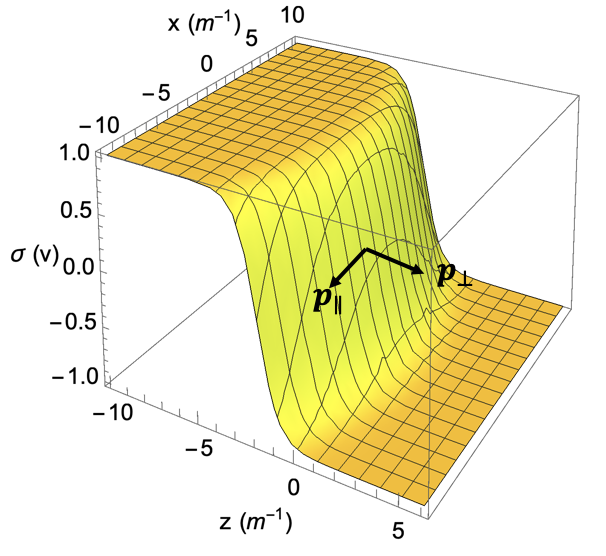}
	\caption{The illustration of DW configuration. }\label{fig:DW3D}
\end{figure}

A typical DW configuration is shown in Fig.~\ref{fig:DW3D}, where we only present the $x$ and $z$ coordinates of the space. We can decompose the comoving momenta $\bp$ and $\bq$ in Eq.~(\ref{eq:ppqq}) into the direction that is in perpendicular to the surface of the wall (the $z$ direction as in Fig.~\ref{fig:DW3D}), and the directions that in parallel to the wall surface. For a generic DW, if we neglect its thickness it can be seen as a Heaviside theta function. In our case, if we only consider the field configuration along the $z$ direction, it can be written as
\bea
\sigma(z) = v(1 - 2\Theta(z)) \ .
\eea
The Fourier transformation of the above configuration is 
\bea
\ts(p_z) = -\frac{i v}{\pi p_z} \ .
\eea
Therefore, we expect that, for a general configuration, $\ts(\bp)$ has the form
\bea
\ts(\bp,\tau) = \frac{\tilde \sigma_\parallel (\bp_\parallel,\tau)}{p_\perp} \ 
\eea
in the region that $\mkz a_c < p < d_w a(\tau)$, where $d_w$ is the physical wall thickness. As an example, in the Landau-Ginzburg type model, as discussed in this work, the DW configuration can be approximated as 
\bea
\sigma(z) = v \tanh\left(\frac{mz}{\sqrt{2}}\right) \ .
\eea
The Fourier transformation is
\bea
\ts(p_z) = \frac{i v}{2^{1/2}m} {\rm csch}\left( \frac{\pi p_z}{\sqrt{2} m} \right) \ .
\eea
Then we know that for $p_z < m$, $\ts(p_z)$ drops as $p_z^{-1}$, and for $p_z > m$, $\ts(p_z)$ drops exponentially.

In the parallel directions, since the oscillations on the surface of the wall are fast diluted by the expansion of the universe, the configuration changes slowly. Therefore we don't expect $\ts_\parallel$ to have significant support for $p_\parallel > \mkz a_c^{-1}$. Now, we can decompose the $\bp$ and $\bq$ integrals in Eq.~(\ref{eq:ppqq}) into $d^2 p_\parallel d p_\perp d^2 q_\parallel d q_\perp$, and then it becomes
\bea\label{eq:pq}
&&\int\frac{d^2\bp_\parallel dp_\perp}{(2\pi)^3} \int\frac{d^2\bq_\parallel dq_\perp}{(2\pi)^3} \left[ \frac{2}{5}(\bp_\parallel\cdot\bq_\parallel + p_\perp q_\perp)^2  \right. \nnn
&& ~~~~\left.- \frac{2}{15} (p_\parallel^2 + p_\perp^2) (q_\parallel^2 + q_\perp^2) \right] \frac{|\ts_\parallel(\bp_\parallel)|^2}{p_\perp^2} \frac{|\ts_\parallel(\bq_\parallel)|^2}{q_\perp^2}
\eea
where the $p_\perp$ and $q_\perp$ integrals are from $\mkz$ to $d_w^{-1} a(\tau)$. From (\ref{eq:pq}) one can see that the $p_\perp$ and $q_\perp$ integrals are dominated in the region $p_\perp \sim d_w^{-1} a(\tau)$ and $q_\perp \sim d_w^{-1} a(\tau)$, whereas the supports of the $p_\parallel$ and $q_\parallel$ integrals are around $\mkz a_c$. Thus, when estimate the value of (\ref{eq:pq}), we can use the approximation $p_\parallel, q_\parallel \ll p_\perp, q_\perp$. Another thing to notice is that $\ts$ must be proportional to $v_{\rm temp}$, the temporary vev of the $\sigma$ field. Then (\ref{eq:pq}) can be estimated as 
\bea\label{eq:A4}
{\cal A}_4 \times \frac{d_w^{-2} a^2(\tau) v_{\rm temp}^4}{[\mkz a_c]^4} \ ,
\eea
where $[\mkz a_c^{-1}]^4$ in the denominator is from dimensional analysis and ${\cal A}_4$ collects all the numerical factors when calculating the integral. 

\begin{figure}[htpb]
	\centering
	\includegraphics[width=0.9\linewidth]{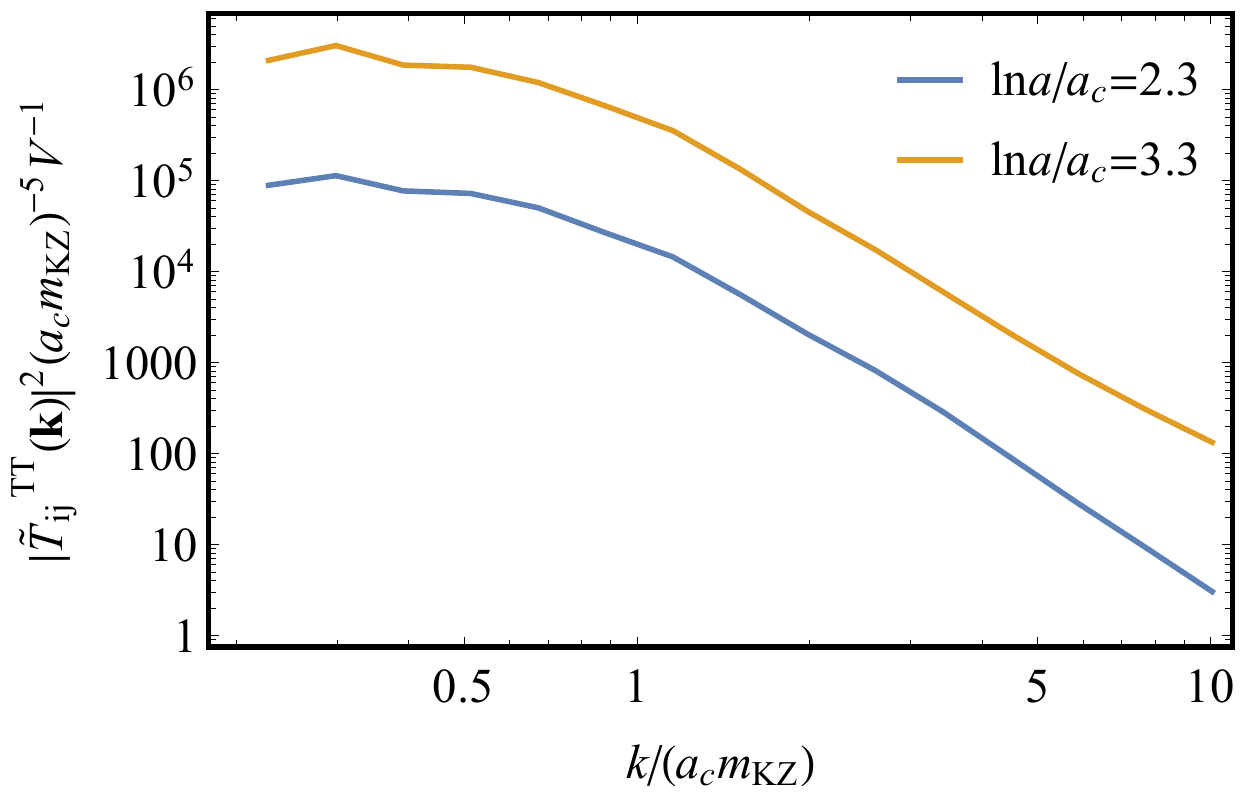}
	\caption{$|\tT_{ij}|^2$ at different time as a function of $k$. }\label{fig:tk}
\end{figure}

Fig.~\ref{fig:tk} shows the $|\tT_{ij}|^2$ at different time after the phase transition. Once can see that the distributions are flat for small $k$, and drops as $k^{-3} \sim k^{-4}$ when $k > \mkz a_c$.
%
%

Together with (\ref{eq:hhff}), (\ref{eq:A3}), (\ref{eq:A4}), Eq.~(\ref{eq:rhoGW}) in the main text, and $a(\tau = -2/k) = k/(2H)$ , we arrive at 
\bea
\frac{d\rho_{\rm GW}}{d\ln k} = {\cal A}_1 \frac{G_N |{\cal E}^i_0|^2 k^3}{H^2 (\mkz a_c) a^4} \frac{v_{\rm temp}^4}{d_w^2} \ ,
\eea
where ${\cal A}_1$ collects all the numerical factors. Now, let's focus on the case that reheating happens fast and finish within one e-fold. Then, as shown in \cite{An:2022cce}, 
\bea
{\cal E}^i_0 = \frac{a_{\rm end}^2 H}{k} \ ,
\eea
where $a_{\rm end}$ is the scale factor at the end of inflation. Thus we have, in the RD era,
\bea
\frac{d\rho_{\rm GW}}{d\ln k} = {\cal A}_1 \frac{G_N  k }{\mkz a_c} \frac{v_{\rm temp}^4}{d_w^2} \left(  \frac{a_{\rm end}}{a}\right)^4 \ ,
\eea
In the instantaneous reheating case, neglecting the change of the number of relativistic degrees of freedom, the radiation energy density can be written as
\bea
\rho_R = \frac{3 H^2}{8\pi G_N} \left( \frac{a_{\rm end}}{a} \right)^4 \ .
\eea
Thus, we have
\bea
\Omega_{\rm GW} = \Omega_R {\cal A}_1 \frac{8\pi G_N^2 k}{3 H^2 \mkz a_c} \frac{v_{\rm temp}^4}{d_w^2} \ .
\eea
In Landau-Ginzburg type model, as shown in Eq.~(\ref{eq:1}) we have 
\bea
d_w \sim m_{\rm temp}^{-1}\ , \;\; \Delta\rho = \frac{m_{\rm temp}^4}{\lambda} \ , \;\; v_{\rm temp}^2 = \frac{m_{\rm temp}^2}{\lambda} \ .
\eea
Therefore we have
\bea
\Omega_{\rm GW} = \Omega_R {\cal A}_1 \frac{8\pi G_N^2 k}{3 H^2 \mkz a_c} \Delta\rho^2 d_w^2 \ .
\eea
During inflation, we have 
\bea
H^2 = \frac{8\pi G_N \rho}{3} \ .
\eea
Then we have
\bea\label{eq:GWpeak2}
\Omega_{\rm GW} = \Omega_R {\cal A}_1 \frac{3}{8\pi} \frac{k}{\mkz a_c}\left(\frac{\Delta\rho}{\rho}\right)^2 H^2 d_w^2 \ .
\eea
This result assumes $k \ll m a_c$ and thus we can neglect the $k$ dependence in $\ts$ when calculating $|\tT_{ij}|^2$. However, we see in Fig.~\ref{fig:tk}, $|\tT_{ij}|^2$ drops significantly when $k > \mkz a_c$. Therefore the peak position of $\Omega_{\rm GW}$ is around $k = \mkz a_c$, and we arrive at the formula
\bea\label{eq:GWpeak}
\Omega_{\rm GW}^{\rm peak} = \Omega_R {\cal A} (H d_w)^2 \left(\frac{\Delta \rho}{\rho_{\rm inf}} \right)^2 \ ,
\eea
where ${\cal A}$ collects all the numerical factors. 

For scenario (A), as shown in Fig.~\ref{fig:sigma} in the main text, $v_{\rm temp}$ arrives at $m\lambda^{-1/2}$ within about one e-fold after the critical time. Thus, to estimate $\Omega_{\rm GW}$, we can just use the parameters in the Lagrangian. However, for scenario (B), $v_{\rm temp}$ changes slowly with $\phi_0$ after the phase transition. Therefore, to have an empirical formula for $\Omega_{\rm GW}^{\rm peak}$ for scenario (B), we take $\tau = - 2/(\mkz a_c)$ to evaluate $d_w$ and $\Delta\rho$.  

\begin{figure}[htpb]
	\centering
	\includegraphics[width=0.9\linewidth]{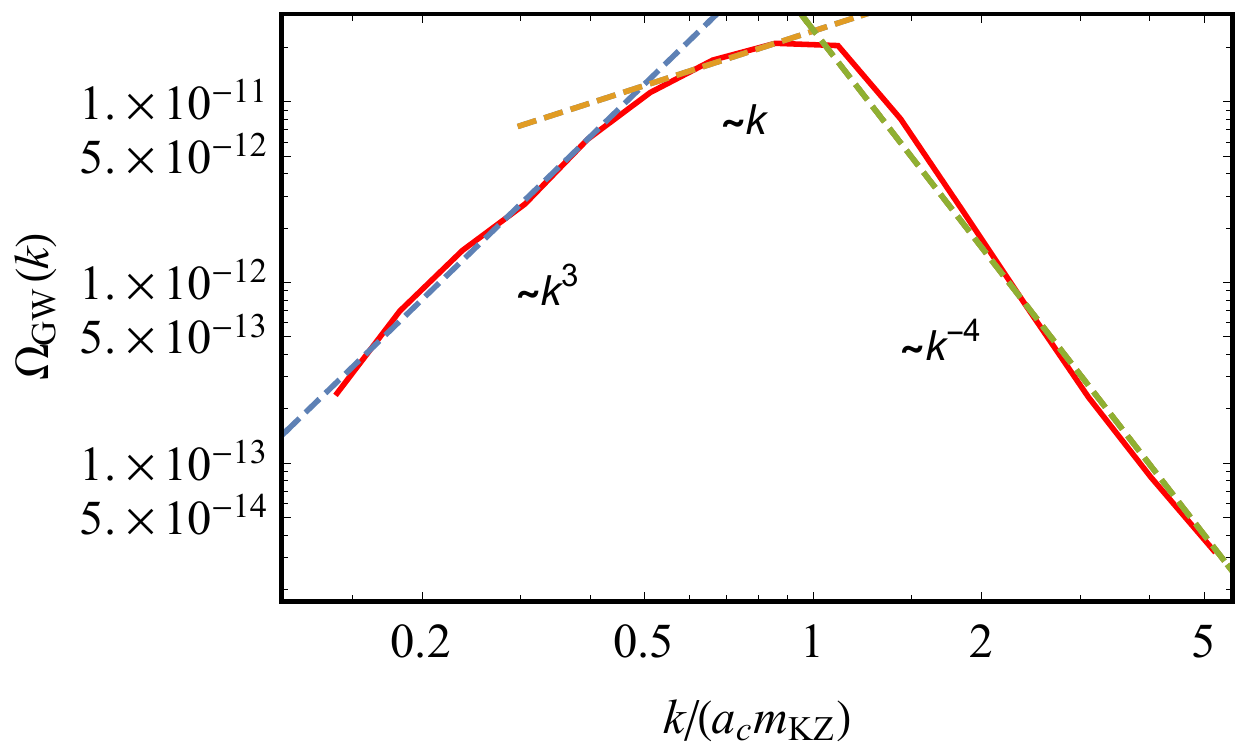}
	\caption{Qualitative behavior of the GW spectrum. Here we assume that the modes reenter the horizon during RD era. In the far IR region, $\Omega_{\rm GW}$ increases as $k^3$, and then transits into $k^1$ when approaching the peak. Then it decays as $k^{-4}$ in the UV region. }\label{fig:peakfit}
\end{figure}

A by-product of (\ref{eq:GWpeak2}) is that it predicts that on the left side of the peak, $\Omega_{\rm GW}$ increases as $k^1$ until it reaches the peak. For $k \ll \mkz a_c$, from causality argument, we have $\Omega_{\rm GW} \sim k^3$~\cite{Cai:2019cdl}. Thus Eq.~(\ref{eq:GWpeak2}) predicts that there is a transition of $\Omega_{\rm GW}$ from $k^3$ to $k^1$ before reaching the peak frequency. This behavior is clearly shown by the numerical simulation presented in Fig.~\ref{fig:peakfit}. The $k^1$ behavior before reaching the peak can be seen as a nontrivial check to the numerical simulation. 

\begin{figure}[htpb]
	\centering
	\includegraphics[width=0.9\linewidth]{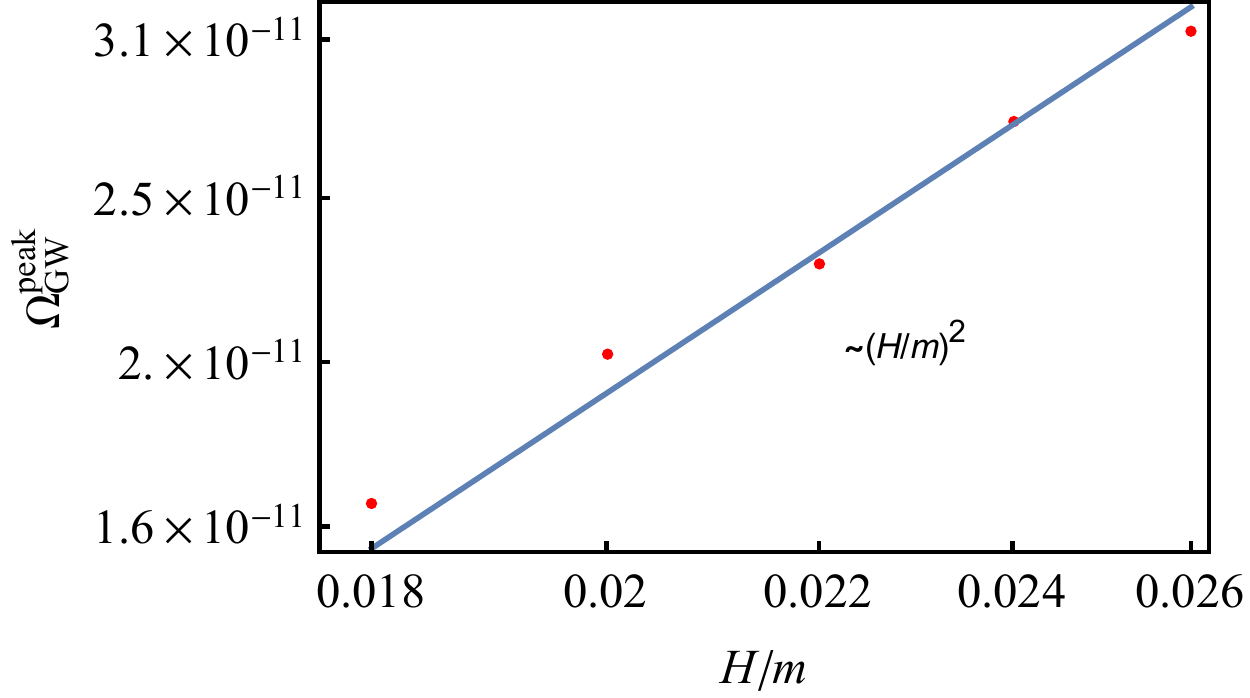}
	\caption{Qualitative behavior of the GW spectrum. We plot $\Omega^{\rm peak}_{\rm GW}$ as a function of $H/m$.  }\label{fig:GWH}
\end{figure}

The analytical formula (\ref{eq:GWpeak}) also shows that $\Omega^{\rm peak}_{\rm GW}$ is proportional to $H^2$. This is also confirmed by the numerical simulations. In Fig.~\ref{fig:GWH}, we show the $\Omega_{\rm GW}^{\rm peak}$ as a function $H/m$. We can see clearly that the qualitative behavior shown in (\ref{eq:GWpeak}) is correct.

%

\section{Evolution of the universe between inflation and reheating}

After inflation, the universe might undergo an intermediate stage before going into the RD era. The GW modes may reenter the horizon in the intermediate stage, and then the spectrum will have a distinctive spectrum. In general, in such a stage the scale factor can be parameterized as $a(t) \sim t^p$ with $p < 1$. Using conformal time the scale factor of the intermediate stage can be written as
\begin{equation}
	a=	a_{\text{end}}\left(\frac{\tau}{\tau_{\text{end}}}\right)^{\frac{p}{1-p}},
\end{equation}
where the subscript ``end'' denotes the end of inflation. Then in the intermediate stage $\sh_{ij}$ satisfies   
\begin{equation}
	\tilde{h}''_{ij}\left(\tau,\bk\right)+\frac{p}{1-p}\frac{2}{\tau}\tilde{h}'_{ij}\left(\tau,\bk\right)+k^2 \tilde{h}_{ij}\left(\tau,\bk\right)=0\ .
\end{equation}
For the modes reenter the horizon, the initial conditions can be written as
\begin{equation}
	\tilde{h}_{ij}\left(\tau_{\text{end}},\bk\right)= h^f_{ij}(\bk)\ ,\quad \tilde{h}'_{ij}\left(\tau_{\text{end}},\bk\right) =  0\ .
\end{equation}
The exact solution of this equation can be expressed as
\bea
\tilde{h}_{ij}\left(\tau ,\bk\right) &=&\Gamma[1-\nu] \left(\frac{k\tau }{2}\right)^{\nu} \left[\cos \nu \pi J_{\nu}\left(k\tau \right) \right.  \nnn
&& \left.-\sin \nu \pi Y_{\nu}\left(k\tau \right)\right]  \tilde{h}^f_{ij}(\bk)\ ,
\eea
where $J_\nu$, $Y_\nu$ are the Bessel functions with
\begin{equation}
	\nu=\frac{3}{2}+\frac{1}{p-1}.
\end{equation}

When the universe enters the RD era, $\sh_{ij}$ satisfies 
\begin{equation}
	\tilde{h}''_{ij}\left(\tau ,\bk\right)+\frac{2}{\tau }\tilde{h}'_{ij}\left(\tau ,\bk\right)+k^2 \tilde{h}_{ij}\left(\tau ,\bk\right)=0\ . \nn
\end{equation}
To get the initial conditions we match $\sh$ and $\sh'$ at the transition time between the intermediate stage and the RD era. Then the general form for $\sh_{ij}$ can be written as
\begin{equation}\label{eq:generalRD}
	\tilde{h}_{ij}\left(\tau ,\bk\right)=\frac{A}{k\tau }\left[S\cos\left(k\tau \right)\!+\!C \sin\left(k\tau \right)\right]\tilde{h}^f_{ij}(\bk) \ . \nn
\end{equation}
Here
\begin{equation}
	A=\frac{1}{\frac{1}{2}-\nu}\frac{\Gamma[1-\nu]}{2^{\nu-\frac{1}{2}}\sqrt{\pi}}x^{\frac{1}{2}+\nu},
\end{equation}
\begin{equation}
	\begin{split}
		S=&\cos\left(\frac{x}{\frac{1}{2}-\nu}\right)\sqrt{\frac{\pi}{2}x}\left[\cos\nu\pi J_\nu(x)-\sin\nu\pi Y_\nu(x)\right]\\
		&-\sin\left(\frac{x}{\frac{1}{2}-\nu}\right)\sqrt{\frac{\pi}{2}x}\left[\cos\nu\pi \left(J'_\nu(x)+\frac{1}{2x}J_\nu(x)\right)\right. \\
		&\left.-\sin\nu\pi \left(Y'_\nu(x)+\frac{1}{2x}Y_\nu(x)\right)\right], 
	\end{split}
\end{equation}
\begin{equation}
	\begin{split}
		C=&\sin\left(\frac{x}{\frac{1}{2}-\nu}\right)\sqrt{\frac{\pi}{2}x}\left[\cos\nu\pi J_\nu(x)-\sin\nu\pi Y_\nu(x)\right]\\
		&+\cos\left(\frac{x}{\frac{1}{2}-\nu}\right)\sqrt{\frac{\pi}{2}x}\left[\cos\nu\pi \left(J'_\nu(x)+\frac{1}{2x}J_\nu(x)\right)\right. \\
		&\left.-\sin\nu\pi \left(Y'_\nu(x)+\frac{1}{2x}Y_\nu(x)\right)\right],
	\end{split}
\end{equation}
with
\begin{equation}
	x=\left(\frac{1}{2}-\nu\right)\left(\frac{a_{\text{re}}}{a_{\text{end}}}\right)^{\frac{1}{\frac{1}{2}-\nu}}\frac{k}{a_{\text{end}}H_{\text{inf}}} , 
\end{equation}
where
$a_{\rm re}$ is the scale factor at transition between the intermediate stage and the RD era. The $x$ parameter here can be seen as the ratio between the physical momentum of the mode and the Hubble expansion rate at the transition between the intermediate stage and the RD era. Thus, $x > 1$ means the modes reenter the horizon in the intermediate stage, and $x < 1$ means the modes reenter the horizon after the intermediate stage, in RD.


Then the GW energy density in the RD era can be written as
\bea
\frac{d\rho_{\rm GW}}{d\ln k} = \left(\frac{a_{\text{re}}}{a_{\text{end}}}\right)^{-\frac{1+2\nu}{\frac{1}{2}-\nu}}A^2\left(|S|^2+|C|^2\right) \left.\frac{d\rho_{\rm GW}}{d\ln k}\right|_{\rm IRH} \ ,
\eea
where the subscript ``IRH'' stands for instantaneous reheating. 
On the other hand with nontrivial intermediate stage the evolution of the total energy density of the universe is also different. By simply counting the redshifts, after reheating the energy density of radiation can be written as
\bea
\rho_R = \left(\frac{a_{\text{re}}}{a_{\text{end}}}\right)^{-\frac{1+2\nu}{\frac{1}{2}-\nu}} \left.\rho_R\right|_{\rm IRH} \ .
\eea
Thus we have
\bea
\Omega_{\rm GW}(k) = D(k) \left.\Omega_{\rm GW}(k)\right|_{\rm IRH} \ ,
\eea
where 
\bea
D(k) = A^2\left(|S|^2+|C|^2\right) \ .
\eea
One can see that an intermediate stage can change both the shape and the strength of the GW signal. The distortion factor $D$ satisfies
\begin{align}
	&D(k)=1,\quad x\ll1 \ ,\\
	&D(k)=A^2,\quad x\gg1\ .
\end{align}

\begin{figure}[htpb]
	\centering
	\includegraphics[width=0.9\linewidth]{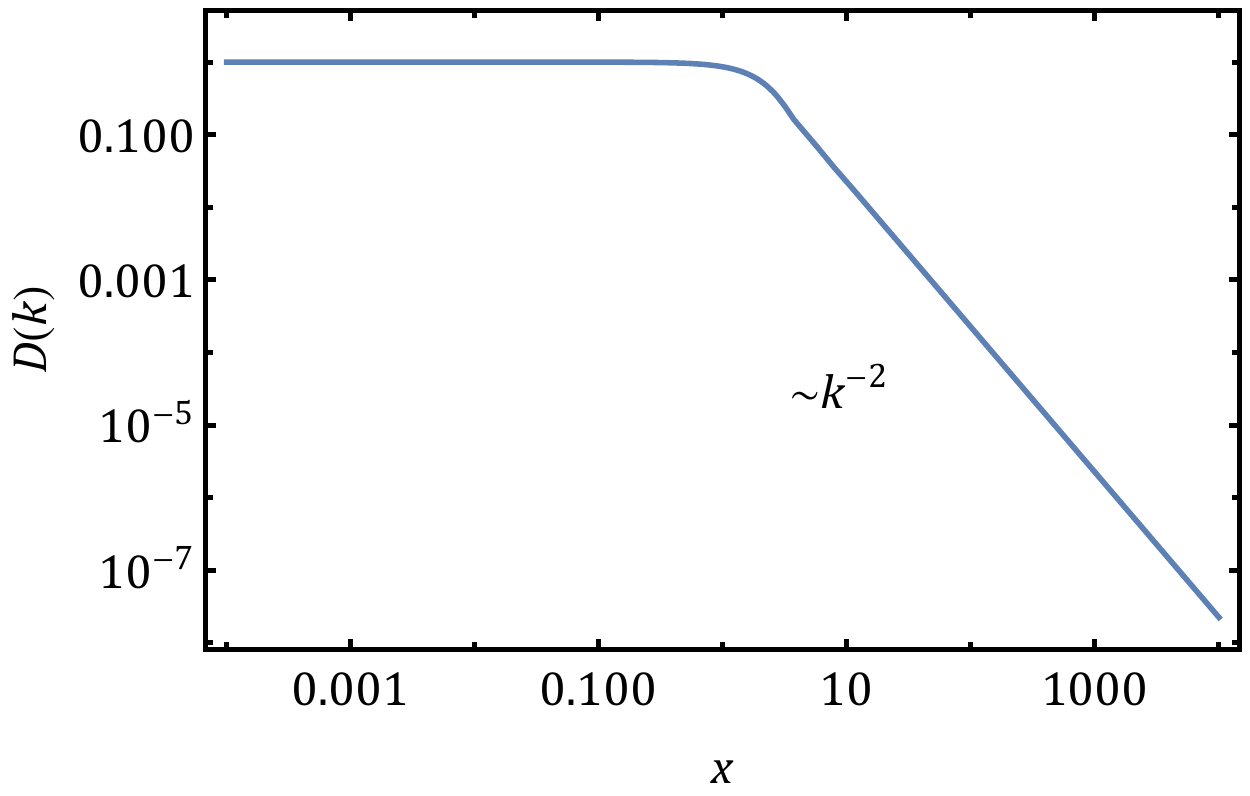}
	\includegraphics[width=0.9\linewidth]{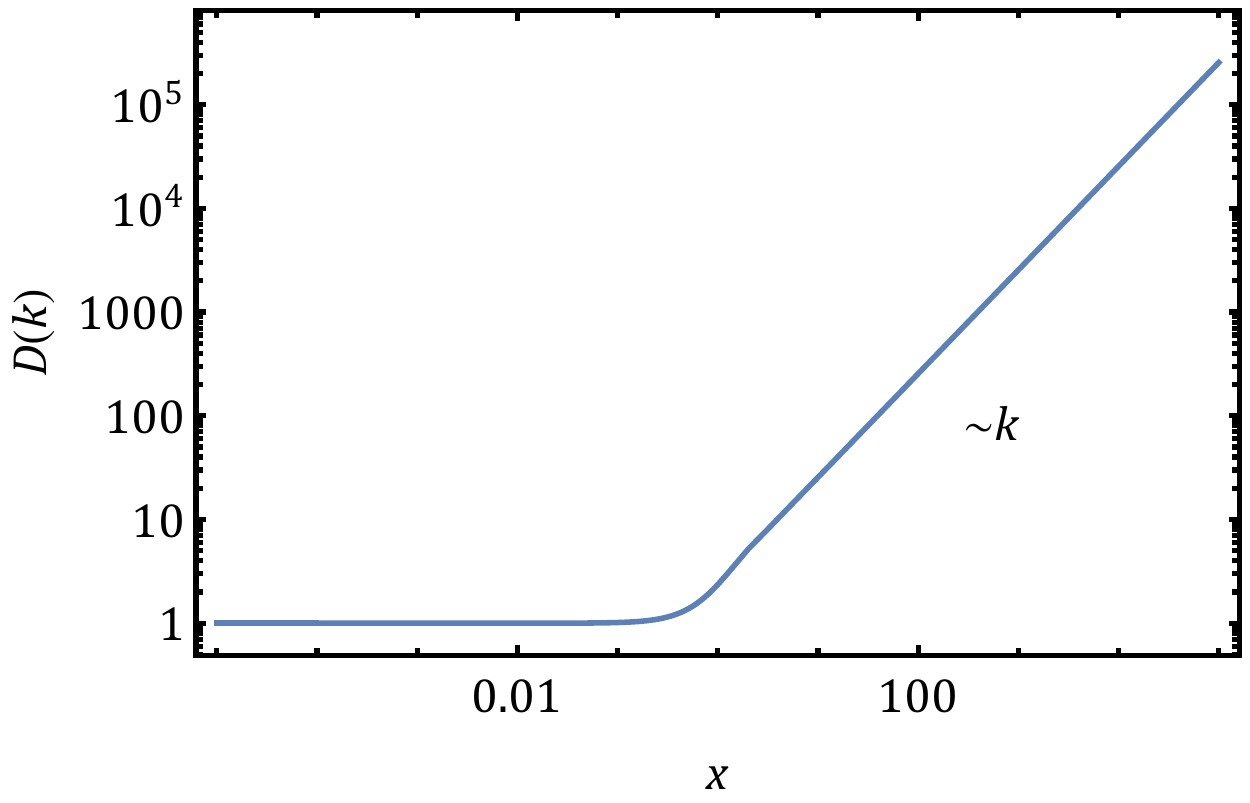}
	\caption{The distortion function $D(k)$ as functions of $x$ for MD intermediate stage (up), and KD intermediate stage (down), respectively. }\label{fig:distort}
\end{figure}

In Fig.~\ref{fig:distort} we show the distortion functions for MD and KD intermediate stage separately. In the $x\gg 1$ limit, which means the mode returns the horizon while the intermediate stage, we have
\begin{align}
	\text{MD:}\ &D(k)=\frac{9}{16}\left(\frac{a_{\text{re}}}{a_{\text{end}}}\right)^{-1}\left(\frac{k}{a_{\text{end}}H_{\text{inf}}}\right)^{-2}=\frac{9}{16} z_{\text{mp}}^{-1}\\
	\text{KD:}\ &D(k)=\frac{4}{\pi}\left(\frac{a_{\text{re}}}{a_{\text{end}}}\right)^2\frac{k}{a_{\text{end}}H_{\text{inf}}}=\frac{4}{\pi}z_{\text{mp}}^{2} \ ,
\end{align}  
where $z_{\rm mp}$ is the redshift from the mode reenters the horizon to the end of the intermediate stage, appeared in Eq.~(\ref{eq:analytical}) in the main text.

\bibliography{ref}
\bibliographystyle{utphys}

\end{document}